\documentclass[reprint,superscriptaddress,amsmath,amssymb,aps]{revtex4-2}
\usepackage{dcolumn}
\usepackage{graphicx}
\usepackage{hyperref}
\usepackage{color}
\usepackage[english]{babel}
\usepackage{bm}% bold math
\usepackage[T1]{fontenc}
\usepackage{multirow}
\usepackage{enumerate}

\usepackage{soul}
\emergencystretch=\maxdimen
\begin{document}
\begin{titlepage}

{\fontsize{20}{10} \textbf{\textcolor{black}{\flushleft{Charge radii of exotic potassium isotopes challenge nuclear theory and the magic character of $N$ = 32}}}}\\

%\\with Forced Linebreak}% Force line breaks with \\
%\thanks{A footnote to the article %title}%

{
\'A.~Koszor\'us$^{1*\dag}$
X.~F.~Yang$^{2,1*}$
W.~G.~Jiang$^{3,4,5}$
S.~J.~Novario$^{3,4}$
S.~W.~Bai$^{2}$
J.~Billowes$^{6}$
C.~L.~Binnersley$^{6}$
M.~L.~Bissell$^{6}$
T.~E.~Cocolios$^{1}$
B.~S.~Cooper$^{6}$
R.~P.~de~Groote$^{7,8}$
A.~Ekstr{\"o}m$^{5}$
K.~T.~Flanagan$^{6,9}$
C.~Forss\'{e}n$^{5}$
S.~Franchoo$^{10}$
R.~F.~Garcia Ruiz$^{11,12}$
F.~P.~Gustafsson$^{1}$
G.~Hagen$^{4,3}$
G.~R.~Jansen$^{4}$
A.~Kanellakopoulos$^{1}$
M.~Kortelainen$^{7,8}$
W.~Nazarewicz$^{13}$
G.~Neyens$^{1,11}$
T.~Papenbrock$^{3,4}$
P.-G.~Reinhard$^{14}$
B.~K.~Sahoo$^{15}$
C.~M.~Ricketts$^{6}$
A.~R.~Vernon$^{1,6}$
S.~G.~Wilkins$^{16}$
}

{
\fontsize{7}{10}{
\selectfont
$^{1}$KU Leuven, Instituut voor Kern- en Stralingsfysica, B-3001 Leuven, Belgium.
$^{2}$School of Physics and State Key Laboratory of Nuclear Physics and Technology, Peking  University, Beijing 100871, China.
$^{3}$Department of Physics and Astronomy, University of Tennessee, Knoxville, Tennessee 37996, USA.
$^{4}$Physics  Division,  Oak  Ridge  National  Laboratory,  Oak  Ridge,  TN  37831,  USA.
$^{5}$Department of Physics, Chalmers University of Technology, SE-412 96 G{\"o}teborg, Sweden.
$^{6}$School of Physics and Astronomy, The University of Manchester, Manchester M13 9PL, United Kingdom.
$^{7}$Department of Physics, University of Jyv\"askyl\"a, PB 35(YFL) FIN-40351 Jyv\"askyl\"a, Finland.
$^{8}$Helsinki Institute of Physics,University of Helsinki, P.O. Box 64, FI-00014 Helsinki, Finland.
$^{9}$Photon Science Institute Alan Turing Building, University of Manchester, Manchester M13 9PY, United Kingdom.
$^{10}$Institut de Physique Nucl\'{e}aire Orsay, IN2P3/CNRS, 91405 Orsay Cedex, France.
$^{11}$Experimental Physics Department, CERN, CH-1211 Geneva 23, Switzerland.
$^{12}$Massachusetts Institute of Technology, Cambridge, MA 02139, USA.
$^{13}$Department of Physics and Astronomy and FRIB Laboratory.
Michigan State University, East Lansing, Michigan 48824, USA.
$^{14}$Institut f{\"u}r Theoretische Physik, Universit{\"a}t Erlangen, Erlangen, Germany.
$^{15}$Atomic, Molecular and Optical Physics Division, Physical Research Laboratory, Navrangpura, Ahmedabad 380009, India.
$^{16}$Engineering Department, CERN, CH-1211 Geneva 23, Switzerland.
%$^{*}$ These authors contributed equally to this work.
$^{*}$ Corresponding authors: Xiaofei Yang (xiaofei.yang@pku.edu.cn), \'A.~Koszor\'us (A.Koszorus@liverpool.ac.uk)
$^{\dag}$ Present address: Oliver Lodge Laboratory, University of Liverpool, Liverpool, UK.
}
}

\vspace{0.1cm}
\end{titlepage}

\textbf{
Nuclear charge radii are sensitive probes of different aspects of the nucleon-nucleon interaction and the bulk properties of nuclear matter; thus, they provide a stringent test and challenge for nuclear theory. The calcium region has been of particular interest, as experimental evidence has suggested a new \lq magic' number at $\textbf{\textit{N =}}$ 32 \cite{Huck1985, Wienholtz2013, ISOLTRAP_K}, while the unexpectedly large increases in the charge radii \cite{GarciaRuiz2016, KREIM201497} open new questions about the evolution of nuclear size in neutron-rich systems.
By combining the collinear resonance ionization spectroscopy method with $\beta$-decay detection, we were able to extend the charge radii measurement of potassium \mbox{($\textbf{\textit{Z =}}$ 19)} isotopes up to the exotic $^{52}$K \mbox{($t_{1/2}$ = 110 ms)}, produced in minute quantities. Our work provides the first charge radii measurement beyond \mbox{$\textbf{\textit{N =}}$ 32} in the region, revealing no signature of the magic character at this neutron number. The results are interpreted with two state-of-the-art nuclear theories. For the first time, a long sequence of isotopes could be calculated with coupled-cluster calculations based on newly developed nuclear interactions. The strong increase in the charge radii beyond \mbox{$\textbf{\textit{N =}}$ 28} is not well captured by these calculations, but is well reproduced by Fayans nuclear density functional theory, which, however, overestimates the odd-even staggering effect. These findings highlight our limited understanding on the nuclear size of neutron-rich systems, and expose pressing problems that are present in some of the best current models of nuclear theory.}

\begin{figure*}[!t]
\includegraphics[width=1.0\textwidth]{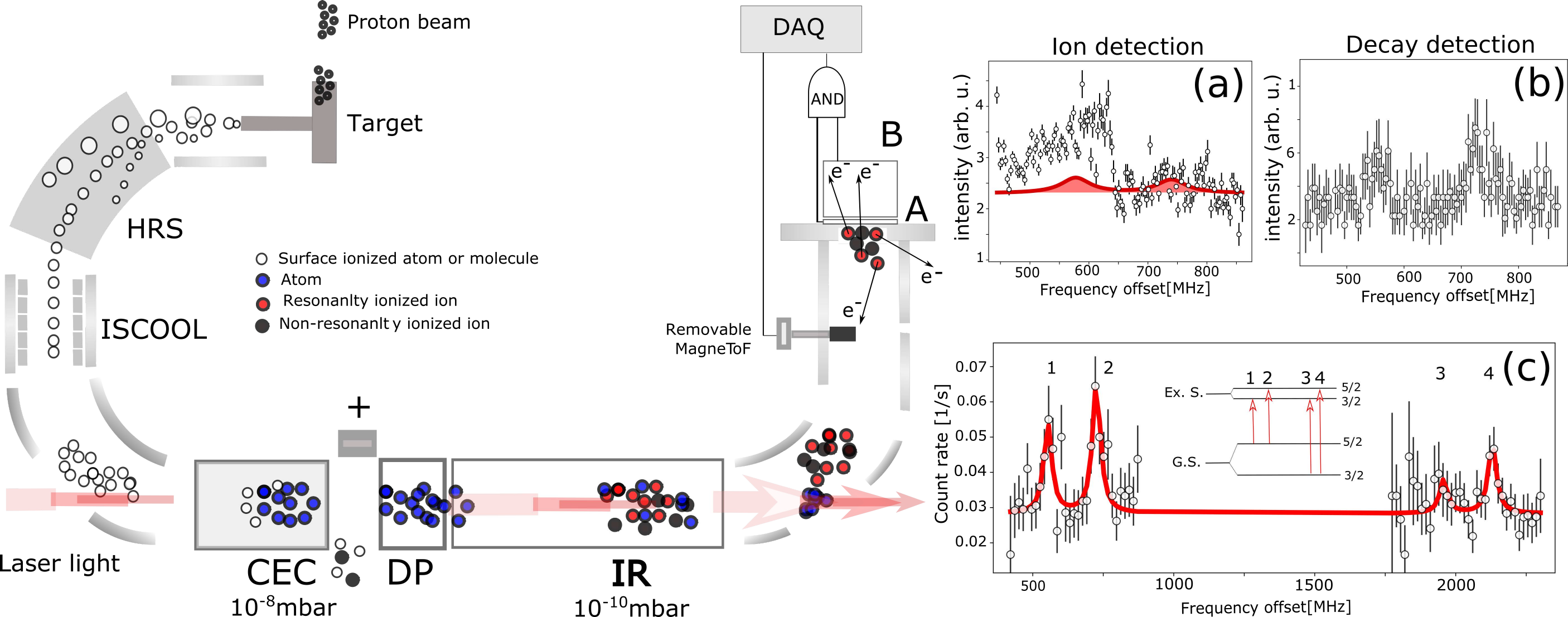}
\vspace{-3mm}
\caption{Left: The CRIS setup at ISOLDE, CERN. The nuclei of interest were
  produced via various nuclear reactions after a 1.4-GeV proton beam
  impinged onto a UCx target. These diffused out of the target, into an
  ion source, and underwent surface ionization. The ion beam was then
  mass separated using a High Resolution Separator
  (HRS), and subsequently cooled and bunched in a linear
  Paul trap (ISCOOL). The bunched ion beam was guided towards the
  CRIS beamline, where the ions were first neutralized in a
  charge-exchange cell filled with potassium vapor. The neutral
  atoms were then delivered to the interaction region. Here the
  bunched beam of atoms was collinearly overlapped with the laser pulses to
  achieve resonance laser ionization. The ionized radioactive
  potassium ions could then be detected using either a MagneToF ion
  detector shown in (a), or plastic scintillator detectors
  (b). (c) The hfs of $^{52}$K measured with the scintillator detectors. Figures (a), (b) and (c) show the detected events as a function of the laser frequency detuning.}
\label{CRIS}
\vspace{-3mm}
\end{figure*}

The charge radius is a fundamental property of the atomic nucleus. Though it globally scales with the nuclear mass as $A^{1/3}$, the nuclear charge radius additionally exhibits appreciable isotopic variations that are the result of complex interactions between protons and neutrons. Indeed, charge radii reflect various nuclear structure phenomena such as halo structures~\cite{Be_halo,He_halo}, shape staggering~\cite{Marsh2018} and shape coexistence~\cite{Zn_yang}, pairing correlations \cite{Curadii,Miller2019}, neutron skins\cite{Hagen2016} and the occurrence of nuclear magic numbers \cite{KREIM201497, Gorges2019,Pb-region}. The term `magic number' refers to the number of protons or neutrons corresponding to completely filled shells, thus resulting in an enhanced stability and a relatively small charge radius.

In the nuclear mass region near potassium, the isotopes with neutron number $N =32$ are proposed to be \lq magic', based on an observed sudden decrease in binding energy beyond $N=32$ \cite{ISOLTRAP_K,Wienholtz2013} and high excitation energy of the first excited state in $^{52}$Ca \cite{Huck1985}. The nuclear charge radius is also a sensitive indicator of \lq magicity': a sudden increase in the charge radii of isotopes is observed after crossing the classical neutron magic numbers $N=$ 28, 50, 82 and 126 \cite{KREIM201497,GaRaii,Gorges2019,Pb-region}. Therefore, the experimentally observed strong increase in the charge radii between $N = 28$ and $N = 32$ of calcium \cite{GarciaRuiz2016} and potassium chains \cite{KREIM201497}, and in particular the large radius of $^{51}$K and $^{52}$Ca (both having 32 neutrons) have attracted significant attention.

One aim of the present study is therefore to shed light on several open questions in this region: how does the nuclear size of very neutron-rich nuclei evolve, and is there any evidence for the \lq magicity' of \mbox{$N = 32$} from nuclear size measurement? We furthermore provide new data to test several newly developed nuclear models, which aim at understanding the evolution of nuclear charge radii of exotic isotopes with large neutron-to-proton imbalance. So far, all $\textit{ab initio}$ nuclear methods, allowing for systematically improvable calculations based on realistic Hamiltonians with nucleon-nucleon and three-nucleon potentials, have failed to explain the enhanced nuclear sizes beyond $N=28$ in the calcium isotopes~\cite{GarciaRuiz2016, SomaPRC}. Meanwhile, nuclear Density Functional Theory (DFT) using Fayans functionals has been successful in predicting the increase in the charge radii of isotopes in the proton-magic calcium chain \cite{Miller2019}, as well as the kinks in the proton-magic tin and lead \cite{Gorges2019}. These state-of-the-art theoretical approaches have been predominantly used to study the charge radii of \mbox{even-$Z$} isotopes, and only very recently could charge radii of \mbox{odd-$Z$} Cu ($Z = 29$) isotopes be investigated with A-body in-medium similarity renormalization group (IM-SRG) method and Fayans-DFT \cite{Curadii}.

Laser spectroscopy techniques yield the most accurate and precise measurements of the charge radius for radioactive nuclei. These highly efficient and sensitive experiments at radioactive ion beam facilities have expanded our knowledge of nuclear charge radii distributed throughout the nuclear landscape \cite{CAMPBELL2016127}. Laser spectroscopy achieves this in a  nuclear-model-independent way by measuring the small perturbations of the atomic hyperfine energy levels due to the electromagnetic properties of the nucleus. Although these hyperfine structure (hfs) effects are as small as one part in a million compared to the total transition frequency, they can nowadays be measured with remarkable precision and efficiency, even for short-lived, weakly-produced, exotic isotopes \cite{Curadii}.

To enhance the sensitivity of the high-resolution, optically detected collinear laser spectroscopy method that was previously used to measure the mean-square charge radii of the potassium isotopes \cite{TOUCHARD1982169,Rossi,KREIM201497}, we used the  Collinear Resonance Ionization Spectroscopy (CRIS) experimental setup at the ISOLDE facility of CERN. This allows very exotic isotopes to be studied with the same resolution as the optically detected method \cite{Curadii,RubenPRL}. Relevant details of the ISOLDE radioactive beam facility and the CRIS setup are depicted in \mbox{Fig.~\ref{CRIS}} (see Methods for details). The CRIS method relies on the step-wise resonant laser excitation and ionization of atoms. For this experiment, a narrowband laser was used to excite potassium atoms from one of the hfs components of the atomic ground state into a hyperfine energy level of an excited state. From there, another excitation and subsequent laser ionization were induced by broadband high-power laser beams, as discussed in Ref. \cite{AGIPRC2019}. The resulting ions were deflected from the remaining (neutral) particles in the beam, and were detected with an ion detector. This method allows nearly background-free ion detection, and thus has very high sensitivity, provided that the contaminating beam particles are not ionized (through e.g. collisional ionisation or due to the high-power non-resonant laser beams). By counting the ions as a function of the laser frequency, the energy differences between the atomic hyperfine transitions were measured. If measurements are performed on more than one isotope, the difference in mean-square charge radius of these isotopes can be obtained from the difference in the hfs
centroid frequency of two isotopes (the isotope shift) with mass numbers $A$ and $A'$:
$\delta \nu^{A,A'} = \nu^{A} - \nu^{A'}$.

In order to apply the CRIS method to study a light element such as potassium, where the optical transition exhibits a lower sensitivity to the nuclear properties, the long-term stability and accurate measurement of the laser frequency had to be investigated. The details of the relevant developments are presented in Ref. \cite{KOSZORUS2019}, where the method was validated by measuring the mean-square charge radii of $^{38-47}$K isotopes with high precision. For the most exotic isotope, there was an additional challenge: a large isobaric contamination at mass $A=52$, measured to be 2$\times$10$^4$ times more intense than the $^{52}$K beam of interest. The resulting detected background rate was found to be an order of magnitude higher than that of the resonantly ionized $^{52}$K ions. In addition, this background rate was found to strongly fluctuate in time, making a measurement with ion detection impossible (see the  hfs spectrum in \mbox{Fig. \ref{CRIS}}(a)).
Taking advantage of the short half-life of $^{52}$K \mbox{($t_{1/2}$ = 110 ms)} and the fact that the isobaric contamination is largely due to the stable isotope $^{52}$Cr, an alternative detection setup was developed, which can distinguish the stable contamination from the radioactive $^{52}$K. A thin and a thick scintillator detector were installed behind the CRIS setup (\mbox{Fig. \ref{CRIS}}). These detectors were used to count the $\beta$-particles emitted though the decay of $^{52}$K. With this setup, the fluctuations in the background rate and signal-to-background ratio were significantly improved, as seen in \mbox{Fig. \ref{CRIS}(b)}. The obtained hfs spectrum of $^{52}$K is presented in \mbox{Fig. \ref{CRIS}}(c). Note that the hfs spectra of $^{47-51}$K were re-measured with the standard CRIS method, and $^{50,51}$K were measured with both ion- and $\beta$-detection. This allows a consistent calculation of isotope shifts of $^{47-52}$K (See Tab~ \ref{table_evaluation} in Methods for details).

The changes in the mean-square charge radii $\delta \langle
r^2\rangle$ are calculated from the isotope shift $\nu^{AA'}$
via
\begin{equation}
    \delta \langle r^2\rangle = \frac{1}{F}\Big[ \nu^{AA'}-(K_{\rm NMS}+K_{\rm SMS}) \frac{m_A-m_{A'}}{(m_A+m_e) m_{A'}}\Big].
\end{equation}
Here $F$, $K_{\rm NMS}$ and $K_{\rm SMS}$ are the atomic field shift,
specific mass shift and normal mass shift factors, respectively (see Methods for
details). Previously published charge radii of potassium isotopes
\cite{TOUCHARD1982169,Behr,KREIM201497,Rossi,AGIPRC2019} have been
extracted from the isotope shifts using an $F$-value calculated with a non-relativistic coupled-cluster method and an empirically determined $K_{\rm SMS}$ value, as reported in Ref.~\cite{Martensson_Pendrill_1990}.

\begin{table}[!t]
\caption{\label{tab:table1} Evaluated experimental isotope shifts $\delta \nu^{39,A}$, differences in mean-square charge radii $\delta\langle r^{2}\rangle$, and charge radii $R_{\rm ch}$ of nuclei $^{36-52}$K. Systematic errors are reported in square brackets. The procedure for the evaluation is discussed in the Methods.}
%\vspace{2mm}
\renewcommand*{\arraystretch}{1.1}
\begin{tabular}{c|cclll}
 $A$& N& $I^{\pi}$ & $\delta \nu^{39,A}$ (MHz)&$\delta\langle r^{2}\rangle^{39,A}$ (fm$^2$)& $R_{\rm ch}$ (fm)\\
\hline
36& 17&2$^{+}$        & -403(9)    &  -0.20(8)[4]    &     3.405(12)[6]  \\
37& 18&3/2$^{+}$      & -264(6)    &  -0.11(6)[3]    &     3.419(8)[4]  \\
38& 19&3$^{+}$        & -126.1(19) &  -0.075(18)[14] &     3.4241(26)[20]  \\
39& 20&3/2$^{+}$      &  0         &  0                 &     3.435(0)[0]  \\
40& 21&4$^{-}$        & 125.63(9)  &  0.025(1)[13]    &     3.4386(1)[19]  \\
41& 22&3/2$^{+}$      & 235.47(9)  &  0.135(1)[26]    &     3.4546(1)[37]  \\
42& 23&2$^{-}$        & 352.4(10)  &  0.128(10)[38]    &     3.4536(15)[55]  \\
43& 24&3/2$^{+}$      & 459.0(12)  &  0.165(11)[49]    &     3.4590(16)[71]  \\
44& 25&2$^{-}$        & 565.1(8)  &  0.163(7)[60]     &     3.4586(11)[87]  \\
45& 26&3/2$^{+}$      & 661.7(16)  &  0.203(15)[70]    &     3.4644(22)[102]  \\
46& 27&2$^{-}$        & 764.1(14)  &  0.150(13)[80]    &     3.4568(19)[116]  \\
47& 28&1/2$^{+}$      & 858.4(9)   &  0.133(8)[90]     &     3.4543(12)[130]  \\
48& 29&1$^{-}$        & 926.4(9)   &  0.328(8)[99]     &     3.4825(12)[142]  \\
49& 30&1/2$^{+}$      & 994.2(11)  &  0.491(10)[108]    &     3.5057(15)[154]  \\
50& 31&0$^{-}$        & 1065.3(10) &  0.592(9)[116]     &     3.5201(13)[165]  \\
51& 32&3/2$^{+}$      & 1130.6(14) &  0.716(13)[124]    &     3.5377(18)[176]  \\
$\bm{52}$&$\bm{33}$ &$\bm{(2^{-})}$& 1198(5) & 0.79(5)[132]           &     3.549(7)[19]  \\
52& 33&(1$^-$)       & 1118(5)     & 1.54(5)[132]      &   3.652(6)[18] \\
52& 33&(3$^-$)       & 1237(5)     & 0.43(5)[132]      &   3.497(7)[19] \\
\hline
\end{tabular}
\end{table}

\begin{figure}[t!]
\centering
\includegraphics[width=0.47\textwidth]{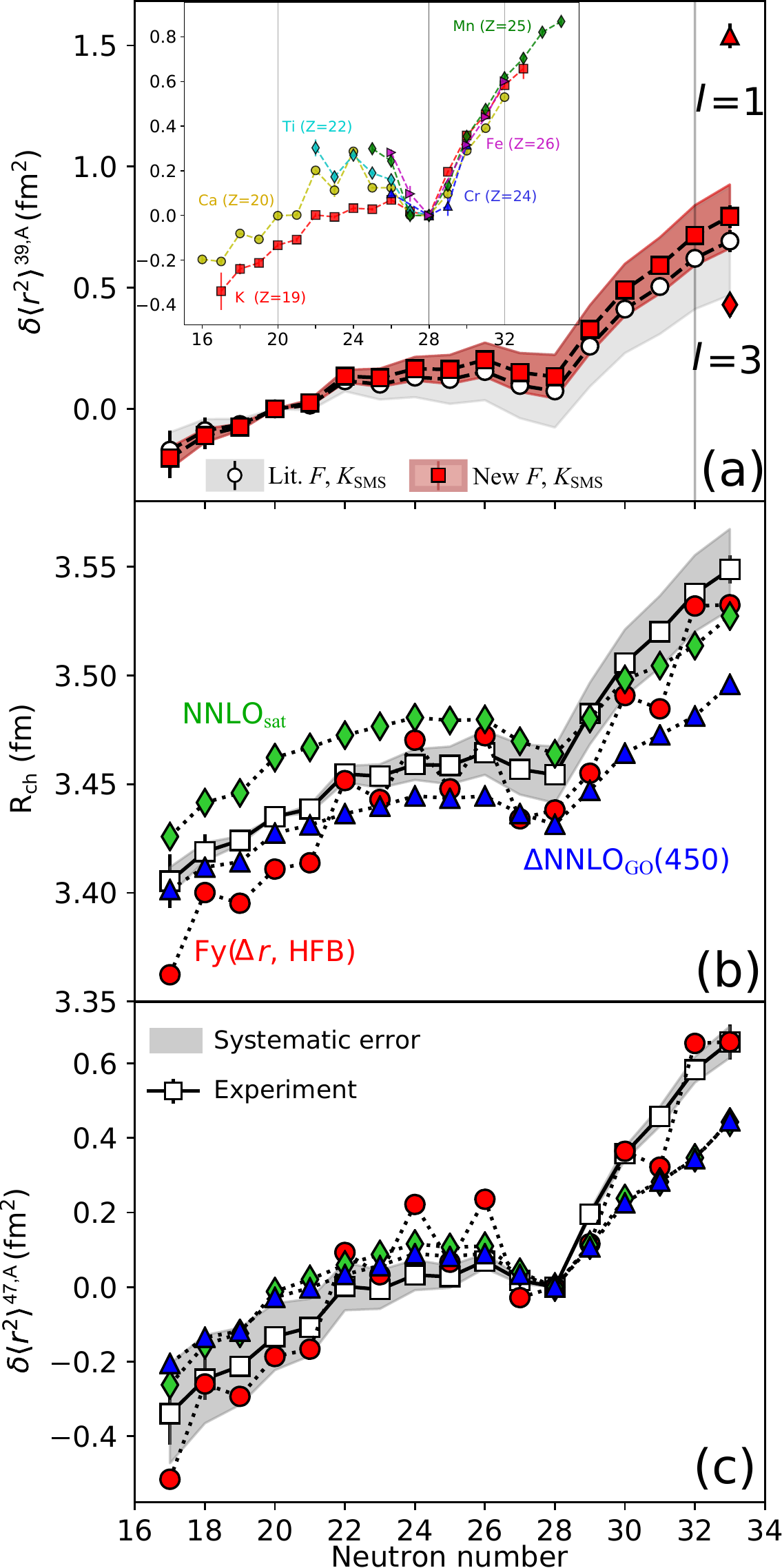}
\vspace{-2mm}
\caption{(a) Changes in the mean-square charge radii of potassium
  isotopes using the newly calculated atomic field shift ($F$) and specific
  mass shift ($K_{\rm SMS}$) factors, as well as the values from
  Ref.~\cite{Martensson_Pendrill_1990}. The red and
  gray bands indicate the uncertainties originating from these atomic
  constants, respectively. The inset shows the changes in the
  mean-square charge radii of neighbouring elements. The agreement of the data for different isotopic chains above $N = 28$ is striking. (b) Comparison of the measured charge radii of
  potassium isotopes with nuclear coupled-cluster calculations
  using two interactions (NNLO$_{\rm sat}$ and  $\Delta$NNLO$_{\rm GO}(450)$) derived from the chiral effective field
  theory, and with the Fayans-DFT calculations with the Fy($\Delta$r,HFB) energy density functional. (c) Changes in the mean-square charge radii of potassium relative to $^{47}$K, in which the systematic uncertainties near $N=28$ are largely cancelled.}
\label{rmsCaK}
\vspace{-3mm}
\end{figure}
We employ the recently developed analytic response relativistic coupled-cluster (ARRCC) theory \cite{Sahoo2020}, an advanced atomic many-body method (see Methods for details), to calculate both the $F$ and $K_{\rm SMS}$ constants. The newly calculated atomic field shift factor, $F = -107.2 (5)$~MHz~fm$^{2}$ is in good agreement with the literature value $F =-110(3)$~MHz~fm$^2$, and is more precise. More importantly, the specific mass shift, a highly correlated atomic parameter, could be calculated from microscopic atomic theory for the first time. The calculated value, $K_{\rm SMS} =- 14.0(22)$~GHz~u, is more precise than the empirical value, $K_{\rm SMS}=- 15.4(38)$~GHz~u from Ref.~\cite{Martensson_Pendrill_1990}, and shows good agreement. Table \ref{tab:table1} presents the isotope shifts, changes in mean-square charge radii, and absolute charge radii of
$^{36-52}$K which were extracted using these new atomic constants. The isotope shifts and charge radii have been re-evaluated using all available data, as described in the Methods section. In Fig.~\ref{rmsCaK}(a) these charge radii are compared with values obtained using the atomic factors taken from
Ref.~\cite{Martensson_Pendrill_1990}. Good agreement is obtained, while the systematic error due to the uncertainty on the atomic factors is clearly reduced. A future measurement of the absolute radii of radioactive potassium isotopes through non-optical means (e.g. electron scattering at the SCRIT facility~\cite{scrit}), would help reduce the systematic uncertainties.

Previously, the nuclear spin and parity of $^{52}$K was tentatively assigned to be $I^\pi=(2^{-})$, based on the very weak feeding into the $^{52}$Ca ground state~\cite{perrot2006}. Here, we have analysed our data assuming two other alternative spin options. Given that the $I=1$ and $I=3$ assumptions produce unrealistically small and large charge radii (see Fig.~\ref{rmsCaK}(a)), our study further supports an $I=2$ assignment.

The inset in Fig.~\ref{rmsCaK}(a) compares the changes in mean-square charge radii (relative to the radius of isotope with neutron number $N=28$) of several isotopic chains in this mass region, up to $Z=26$. A remarkable observation is that the charge radii beyond $N=28$ follow \emph{the same} steep increasing trend, irrespective of the number of protons in the nucleus. Beyond $N=32$, data are only available for potassium (this work) and manganese ($Z=25$) \cite{Hey2016}. Both charge radius trends are very similar, with no signature of a characteristic kink that would indicate \lq magicity' at $N=32$.

%\begin{figure}[!t]
%\includegraphics[width=0.5\textwidth]{New_ch.pdf}
%\caption{The changes in the mean-square charge radii.}
%\label{changes_in_ms_}
%\end{figure}

Previously, $\textit{ab initio}$ coupled cluster (CC) calculations based on the NNLO$_{\rm sat}$ interaction~\cite{AE_sat} were used to describe the nuclear charge radii in calcium isotopes~\cite{Hagen2016,GarciaRuiz2016}. At that time, calculations in this framework could only be performed for spherical isotopes near \lq doubly-magic' nuclei. While these calculations predicted the absolute charge radii near $^{40,48}$Ca very well, they failed to reproduce the observed large charge radii around neutron number $N=32$.  In Fig.~\ref{rmsCaK}(b)  we compare the experimental data to CC calculations that start from a symmetry-breaking reference state, which allows us to compute charge radii of all potassium isotopes (see Methods for details). Results obtained with the NNLO$_{\rm sat}$ interaction significantly overestimate the experimental data near stability, where the experimental uncertainties on the total radii are the smallest. This interaction was fitted to experimental binding energies and charge radii of selected nuclei up to mass number \mbox{$A=25$~\cite{AE_sat}.} Therefore, a newly constructed $\Delta{\rm NNLO}_{\rm GO}(450)$ interaction was developed, which includes pion-physics and effects of the $\Delta(1232)$ isobar. This interaction is constrained by properties of only light nuclei with mass numbers $A \leq 4$ and by nuclear matter at the saturation point (i.e. its saturation energy and density, and its symmetry energy, see Methods for details).
By virtue of including saturation properties, CC calculations using the $\Delta$NNLO$_{\rm GO}(450)$ interaction yield an improvement in the accuracy of the description of potassium charge radii near stability, as shown in Fig.~\ref{rmsCaK}(b). However, both NNLO$_{\rm sat}$ and $\Delta$NNLO$_{\rm GO}(450)$ interactions still underestimate the steep increase observed beyond $N=28$. This is better visualised by plotting the differences in mean-square charge radii relative to $^{47}$K (with neutron number $N = 28$), $\delta \left\langle r^2\right\rangle^{47,A}$, shown in \mbox{Fig. \ref{rmsCaK}(c)}. The systematic uncertainties near $N=28$ are strongly reduced by choosing this reference. It is also worth noting that, for $\delta \left\langle r^2\right\rangle^{47,A}$ below $N=28$, both interactions show a good agreement with experimental results within the systematic uncertainty.  %% Or add inset in Fig 2b.

What can be the reason for the underestimation of charge radii for $N>28$? The reference state in CC calculations is the axial Hartree-Fock state. For nearly spherical nuclei, a general (triaxial) cranked HFB state that breaks time-reversal and gauge symmetries could perhaps provide a better reference. Also, since the angular momentum was not restored, the associated correlations are missing as well in CC results.

The DFT is the method of choice for heavy systems. Nuclei with $Z\approx 20$ cover the region where both methods, DFT and CC, can be successfully applied.
Our DFT calculations use in particular the Fayans functional Fy($\Delta$r,HFB) \cite{Fayan2017} (see Methods for details) which was developed with a focus on charge radii. This method closely reproduces the absolute charge radii of calcium isotopes, including the steep increase beyond $N=28$ \cite{Miller2019}. Furthermore, Fy($\Delta$r,HFB) reproduces the absolute radii of the magic tin \cite{Gorges2019}, cadmium \cite{Hamm18a} as well as the odd-$Z$ copper isotopes \cite{Curadii}. It is to be noted however, that the potassium isotopes are a lighter system in which the polarization effects are expected to be stronger than in the heavier copper isotopes. To account for that, we extended the Fayans-DFT framework to allow for deformed HFB solutions. The isotopic chain contains odd-odd nuclei and the present DFT and CC treatment does not allow a clean spin selection for these. Consequently, the HFB calculations provide an averaged description for the odd-odd isotopes.
As seen in Fig.~\ref{rmsCaK}(b), except for the neutron-poor side, Fy($\Delta$r,HFB) calculations reproduce the average global trend rather well, in particular the steep increase above $N=28$. However, this model grossly overestimates the odd-even staggering. The odd-even staggering in the potassium isotopes is significantly reduced with respect to calcium isotopes. This is very well captured by the CC calculations, due to the fact that these describe in detail the many-body correlations (see Methods for details). In nuclear DFT, local many-body correlations are treated less precisely. %This conclusion is consistent with the findings of Ref. \cite{Curadii}.

In summary, this work presents the first measurement of nuclear charge radii beyond the proposed \lq magic' number $N=32$ in the calcium region. This was achieved by combining collinear resonance ionization spectroscopy with $\beta$-decay detection, enabling the exotic isotope $^{52}$K to be studied, despite its short half-life, low production rate and poor purity. Taking advantage of recent developments in atomic calculations, precise charge radii of the potassium chain were extracted. No sudden change in the $^{52}$K charge radius is observed, thus no signature for \lq magicity' at \mbox{$N=32$} is found. The comparison with nuclear theory predictions for the demanding case of potassium isotopes helps to uncover more about the strengths and open problems in current state-of-the-art nuclear models.
CC calculations based on the new nucleon-nucleon potentials derived from chiral effective field theory, optimized with few-body nuclei properties as well as nuclear saturation properties,
describe very well the absolute nuclear charge radii of the potassium isotopes near stability, and also the small odd-even staggering. However, the steep rise in charge radii above $N = 28$ remains underestimated. The similarity in the performance of $\Delta$NNLO$_\text{GO}(450)$ and NNLO$_\text{sat}$ interactions suggests that the charge radii beyond $N = 28$ are insensitive to the details of chiral interactions at next-to-next-to-leading-order, and some crucial ingredient is lacking in these many-body methods.
The Fayans-DFT model captures the general trend across the measured isotopes and reproduces the absolute radii rather well, including the steep increase up to $N=33$. However, this model overestimates the odd-even staggering significantly. These findings highlight our limited understanding on the size of neutron-rich nuclei, and will undoubtedly trigger further developments in nuclear theory as demanding nuclear data on charge radii keeps uncovering problems with the best current models.
%A future test for the accuracy of the CC and Fayans-DFT methods could be the measurement of the absolute radii of radioactive potassium isotopes through non-optical means (e.g. electron scattering at the SCRIT facility~\cite{scrit}), which would help reduce the systematic uncertainties.
%, e.g. electron scattering at the SCRIT facility~\cite{scrit}.%, or muonic X ray spectroscopy~\cite{muonic}.

%, as was also seen in the Cu isotopes close to the doubly-magic $^{78}$Ni.
%These findings will undoubtedly trigger further developments in nuclear theory. The demanding nuclear data on charge radii, with their global and local isotopic trends, keeps uncovering problems with the best current models.

\vspace{5mm}

\textbf{Acknowledgments} We acknowledge the support of the ISOLDE
collaboration and technical teams. This work was supported in part by
the National Key R\&D Program of China (Contract No: 2018YFA0404403);
the National Natural Science Foundation of China (No:11875073); the BriX
Research Program No. P7/12, FWO-Vlaanderen (Belgium), GOA
15/010 from KU Leuven; ERC Consolidator Grant no. 648381 (FNPMLS); the
STFC consolidated grants ST/L005794/1 and ST/L005786/1 and Ernest
Rutherford Grant No. ST/L002868/1; the EU Horizon2020 research and
innovation programme through ENSAR2 (no. 654002); the U.S. Department
of Energy, Office of Science, Office of Nuclear Physics under grants
DE-00249237, DE-FG02-96ER40963 and DE-SC0018223 (SciDAC-4 NUCLEI collaboration).
This work received funding from the European Research Council (ERC)
under the European Union's Horizon 2020 research and innovation
programme (Grant agreement No. 758027), the Swedish Research Council
grant 2017-04234, and the Swedish Foundation for International
Cooperation in Research and Higher Education (STINT) grant
IG2012-5158. BKS acknowledges use of Vikram-100 HPC cluster of
Physical Research Laboratory, Ahmedabad for atomic calculations.
Computer time was provided by the Innovative and Novel Computational
Impact on Theory and Experiment (INCITE) program. This research used
resources of the Oak Ridge Leadership Computing Facility and of the
Compute and Data Environment for Science (CADES) located at Oak Ridge
National Laboratory, which is supported by the Office of Science of
the Department of Energy under Contract No. DE-AC05-00OR22725.

\textbf{Author contribution}
A.K., X.F.Y., S.W.B. J.B., C.L.B.,
M.L.B., T.E.C.,B.S.C., R.P.d.G,, K.T.F., S.F., R.F.G.R., F.P.G., A.K., G.N., C.M.R,
A.R.V., and S.G.W. performed
the experiment. A.K., X.Y. led the experiment and A.K., X.Y., R.F.G.R. and S.W.B. designed, simulated and installed the $\beta$-detection system. A.K., X.F.Y. performed the data analysis. W.G.J., G.H., T.P., A.E., G.R.J, S.N. C.F. developed the $\Delta$NNLO$_{\rm GO}$ interactions and performed the CC calculation. W.N., P.-G.R., M.K. performed the DFT calculation. B.K.S. performed the atomic physics calculations. A.K., X.F.Y., G.N., W.N., P-G.R., G.H. and T.P. prepared the manuscript. A.K., X.F.Y., G.N., W.N., P.-G.R. and G.H. prepared the figures. All authors discussed the results and contributed to the manuscript at all stages. A.K., X.F.Y contributed equally to this work.

\textbf{Ethics declarations} \\
The authors declare no competing interests.

\textbf{Data Availability Statement}\\
%The data represented in Figs. 1-4 are available as Source Data 1-4.
The data that support the findings of this study are available from the corresponding author upon reasonable request.

\bibliography{apssamp}% Produces the bibliography via BibTeX.

\begin{thebibliography}{10}
\expandafter\ifx\csname url\endcsname\relax
  \def\url#1{\texttt{#1}}\fi
\expandafter\ifx\csname urlprefix\endcsname\relax\def\urlprefix{URL }\fi
\providecommand{\bibinfo}[2]{#2}
\providecommand{\eprint}[2][]{\url{#2}}

\bibitem{Huck1985}
\bibinfo{author}{Huck, A.} \emph{et~al.}
\newblock \bibinfo{title}{Beta decay of the new isotopes $^{52}${K},
  $^{52}${Ca}, and $^{52}${Sc}; a test of the shell model far from stability}.
\newblock \emph{\bibinfo{journal}{Phys. Rev. C}} \textbf{\bibinfo{volume}{31}},
  \bibinfo{pages}{2226--2237} (\bibinfo{year}{1985}).
\newblock \urlprefix\url{https://link.aps.org/doi/10.1103/PhysRevC.31.2226}.

\bibitem{Wienholtz2013}
\bibinfo{author}{Wienholtz, F.} \emph{et~al.}
\newblock \bibinfo{title}{{Masses of exotic calcium isotopes pin down nuclear
  forces}}.
\newblock \emph{\bibinfo{journal}{Nature}} \textbf{\bibinfo{volume}{498}},
  \bibinfo{pages}{346--349} (\bibinfo{year}{2013}).
\newblock \urlprefix\url{http://www.nature.com/articles/nature12226}.

\bibitem{ISOLTRAP_K}
\bibinfo{author}{Rosenbusch, M.} \emph{et~al.}
\newblock \bibinfo{title}{Probing the ${N}=32$ shell closure below the magic
  proton number ${Z}=20$: Mass measurements of the exotic isotopes
  $^{52,53}${K}}.
\newblock \emph{\bibinfo{journal}{Phys. Rev. Lett.}}
  \textbf{\bibinfo{volume}{114}}, \bibinfo{pages}{202501}
  (\bibinfo{year}{2015}).
\newblock
  \urlprefix\url{https://link.aps.org/doi/10.1103/PhysRevLett.114.202501}.

\bibitem{GarciaRuiz2016}
\bibinfo{author}{{Garcia Ruiz}, R.~F.} \emph{et~al.}
\newblock \bibinfo{title}{{Unexpectedly large charge radii of neutron-rich
  calcium isotopes}}.
\newblock \emph{\bibinfo{journal}{Nat. Phys.}} \textbf{\bibinfo{volume}{12}},
  \bibinfo{pages}{594--598} (\bibinfo{year}{2016}).
\newblock \urlprefix\url{http://www.nature.com/articles/nphys3645}.

\bibitem{KREIM201497}
\bibinfo{author}{Kreim, K.} \emph{et~al.}
\newblock \bibinfo{title}{Nuclear charge radii of potassium isotopes beyond
  ${N}=28$}.
\newblock \emph{\bibinfo{journal}{Phys. Lett. B}}
  \textbf{\bibinfo{volume}{731}}, \bibinfo{pages}{97 -- 102}
  (\bibinfo{year}{2014}).
\newblock
  \urlprefix\url{http://www.sciencedirect.com/science/article/pii/S0370269314001038}.

\bibitem{Be_halo}
\bibinfo{author}{N\"ortersh\"auser, W.} \emph{et~al.}
\newblock \bibinfo{title}{{N}uclear {C}harge {R}adii of $^{7,9,10}${Be} and the
  {O}ne-{N}eutron {H}alo {N}ucleus $^{11}${Be}}.
\newblock \emph{\bibinfo{journal}{Phys. Rev. Lett.}}
  \textbf{\bibinfo{volume}{102}}, \bibinfo{pages}{062503}
  (\bibinfo{year}{2009}).
\newblock
  \urlprefix\url{https://link.aps.org/doi/10.1103/PhysRevLett.102.062503}.

\bibitem{He_halo}
\bibinfo{author}{Mueller, P.} \emph{et~al.}
\newblock \bibinfo{title}{Nuclear charge radius of $^{8}\mathrm{He}$}.
\newblock \emph{\bibinfo{journal}{Phys. Rev. Lett.}}
  \textbf{\bibinfo{volume}{99}}, \bibinfo{pages}{252501}
  (\bibinfo{year}{2007}).
\newblock
  \urlprefix\url{https://link.aps.org/doi/10.1103/PhysRevLett.99.252501}.

\bibitem{Marsh2018}
\bibinfo{author}{Marsh, B.~A.} \emph{et~al.}
\newblock \bibinfo{title}{{Characterization of the shape-staggering effect in
  mercury nuclei}}.
\newblock \emph{\bibinfo{journal}{Nat. Phys.}} \textbf{\bibinfo{volume}{14}},
  \bibinfo{pages}{1163--1167} (\bibinfo{year}{2018}).
\newblock \urlprefix\url{http://www.nature.com/articles/s41567-018-0292-8}.

\bibitem{Zn_yang}
\bibinfo{author}{Yang, X.~F.} \emph{et~al.}
\newblock \bibinfo{title}{Isomer shift and magnetic moment of the long-lived
  $1/{2}^{+}$ isomer in $_{30}^{79}${Zn}$_{49}$: Signature of shape coexistence
  near $^{78}{Ni}$}.
\newblock \emph{\bibinfo{journal}{Phys. Rev. Lett.}}
  \textbf{\bibinfo{volume}{116}}, \bibinfo{pages}{182502}
  (\bibinfo{year}{2016}).
\newblock
  \urlprefix\url{https://link.aps.org/doi/10.1103/PhysRevLett.116.182502}.

\bibitem{Curadii}
\bibinfo{author}{de~Groote, R.} \emph{et~al.}
\newblock \bibinfo{title}{Measurement and microscopic description of odd--even
  staggering of charge radii of exotic copper isotopes}.
\newblock \emph{\bibinfo{journal}{Nat. Phys.}} \textbf{\bibinfo{volume}{16}},
  \bibinfo{pages}{620--624} (\bibinfo{year}{2020}).
\newblock \urlprefix\url{https://doi.org/10.1038/s41567-020-0868-y}.

\bibitem{Miller2019}
\bibinfo{author}{Miller, A.~J.} \emph{et~al.}
\newblock \bibinfo{title}{{Proton superfluidity and charge radii in proton-rich
  calcium isotopes}}.
\newblock \emph{\bibinfo{journal}{Nat. Phys.}} \textbf{\bibinfo{volume}{15}},
  \bibinfo{pages}{432--436} (\bibinfo{year}{2019}).
\newblock \urlprefix\url{http://www.nature.com/articles/s41567-019-0416-9}.

\bibitem{Hagen2016}
\bibinfo{author}{Hagen, G.} \emph{et~al.}
\newblock \bibinfo{title}{{Neutron and weak-charge distributions of the
  $^{48}${C}a nucleus}}.
\newblock \emph{\bibinfo{journal}{Nat. Phys.}} \textbf{\bibinfo{volume}{12}},
  \bibinfo{pages}{186--190} (\bibinfo{year}{2016}).
\newblock \urlprefix\url{http://www.nature.com/articles/nphys3529}.

\bibitem{Gorges2019}
\bibinfo{author}{Gorges, C.} \emph{et~al.}
\newblock \bibinfo{title}{{L}aser {S}pectroscopy of {N}eutron-{R}ich {T}in
  {I}sotopes: {A} {D}iscontinuity in {C}harge {R}adii across the ${N}=82$
  {S}hell {C}losure}.
\newblock \emph{\bibinfo{journal}{Phys. Rev. Lett.}}
  \textbf{\bibinfo{volume}{122}}, \bibinfo{pages}{192502}
  (\bibinfo{year}{2019}).
\newblock
  \urlprefix\url{https://link.aps.org/doi/10.1103/PhysRevLett.122.192502}.

\bibitem{Pb-region}
\bibinfo{author}{Goddard, P.~M.}, \bibinfo{author}{Stevenson, P.~D.} \&
  \bibinfo{author}{Rios, A.}
\newblock \bibinfo{title}{Charge radius isotope shift across the ${N}=126$
  shell gap}.
\newblock \emph{\bibinfo{journal}{Phys. Rev. Lett.}}
  \textbf{\bibinfo{volume}{110}}, \bibinfo{pages}{032503}
  (\bibinfo{year}{2013}).
\newblock
  \urlprefix\url{https://link.aps.org/doi/10.1103/PhysRevLett.110.032503}.

\bibitem{GaRaii}
\bibinfo{author}{Procter, T.~J.} \emph{et~al.}
\newblock \bibinfo{title}{Nuclear mean-square charge radii of
  ${^{63,64,66,68}{Ga}}$ nuclei: No anomalous behavior at ${N}=32$}.
\newblock \emph{\bibinfo{journal}{Phys. Rev. C}} \textbf{\bibinfo{volume}{86}},
  \bibinfo{pages}{034329} (\bibinfo{year}{2012}).
\newblock \urlprefix\url{https://link.aps.org/doi/10.1103/PhysRevC.86.034329}.

\bibitem{SomaPRC}
\bibinfo{author}{Som\`a, V.}, \bibinfo{author}{Navr\'atil, P.},
  \bibinfo{author}{Raimondi, F.}, \bibinfo{author}{Barbieri, C.} \&
  \bibinfo{author}{Duguet, T.}
\newblock \bibinfo{title}{Novel chiral hamiltonian and observables in light and
  medium-mass nuclei}.
\newblock \emph{\bibinfo{journal}{Phys. Rev. C}}
  \textbf{\bibinfo{volume}{101}}, \bibinfo{pages}{014318}
  (\bibinfo{year}{2020}).
\newblock \urlprefix\url{https://link.aps.org/doi/10.1103/PhysRevC.101.014318}.

\bibitem{CAMPBELL2016127}
\bibinfo{author}{Campbell, P.}, \bibinfo{author}{Moore, I.} \&
  \bibinfo{author}{Pearson, M.}
\newblock \bibinfo{title}{Laser spectroscopy for nuclear structure physics}.
\newblock \emph{\bibinfo{journal}{Prog. Part. Nucl. Phys.}}
  \textbf{\bibinfo{volume}{86}}, \bibinfo{pages}{127--180}
  (\bibinfo{year}{2016}).
\newblock
  \urlprefix\url{http://www.sciencedirect.com/science/article/pii/S0146641015000915}.

\bibitem{TOUCHARD1982169}
\bibinfo{author}{Touchard, F.} \emph{et~al.}
\newblock \bibinfo{title}{Isotope shifts and hyperfine structure of
  $^{38-47}${K} by laser spectroscopy}.
\newblock \emph{\bibinfo{journal}{Phys. Lett. B}}
  \textbf{\bibinfo{volume}{108}}, \bibinfo{pages}{169 -- 171}
  (\bibinfo{year}{1982}).
\newblock
  \urlprefix\url{http://www.sciencedirect.com/science/article/pii/0370269382911674}.

\bibitem{Rossi}
\bibinfo{author}{Rossi, D.~M.} \emph{et~al.}
\newblock \bibinfo{title}{Charge radii of neutron-deficient $^{36}${K} and
  $^{37}\mathrm{K}$}.
\newblock \emph{\bibinfo{journal}{Phys. Rev. C}} \textbf{\bibinfo{volume}{92}},
  \bibinfo{pages}{014305} (\bibinfo{year}{2015}).
\newblock \urlprefix\url{https://link.aps.org/doi/10.1103/PhysRevC.92.014305}.

\bibitem{RubenPRL}
\bibinfo{author}{de~Groote, R.~P.} \emph{et~al.}
\newblock \bibinfo{title}{{U}se of a {C}ontinuous {W}ave {L}aser and {P}ockels
  {C}ell for {S}ensitive {H}igh-{R}esolution {C}ollinear {R}esonance
  {I}onization {S}pectroscopy}.
\newblock \emph{\bibinfo{journal}{Phys. Rev. Lett.}}
  \textbf{\bibinfo{volume}{115}}, \bibinfo{pages}{132501}
  (\bibinfo{year}{2015}).
\newblock
  \urlprefix\url{https://link.aps.org/doi/10.1103/PhysRevLett.115.132501}.

\bibitem{AGIPRC2019}
\bibinfo{author}{Koszor\'us, A.} \emph{et~al.}
\newblock \bibinfo{title}{Precision measurements of the charge radii of
  potassium isotopes}.
\newblock \emph{\bibinfo{journal}{Phys. Rev. C}}
  \textbf{\bibinfo{volume}{100}}, \bibinfo{pages}{034304}
  (\bibinfo{year}{2019}).
\newblock \urlprefix\url{https://link.aps.org/doi/10.1103/PhysRevC.100.034304}.

\bibitem{KOSZORUS2019}
\bibinfo{author}{Koszor\'us, A.} \emph{et~al.}
\newblock \bibinfo{title}{Resonance ionization schemes for high resolution and
  high efficiency studies of exotic nuclei at the cris experiment}.
\newblock \emph{\bibinfo{journal}{Nucl. Instrum. Methods Phys. Res., Sect. B}}
  (\bibinfo{year}{2019}).
\newblock
  \urlprefix\url{http://www.sciencedirect.com/science/article/pii/S0168583X19302277}.

\bibitem{Behr}
\bibinfo{author}{Behr, J.~A.} \emph{et~al.}
\newblock \bibinfo{title}{Magneto-optic {T}rapping of
  $\ensuremath{\beta}$-{D}ecaying $^{38}${K}$^{\mathit{m}}$, $^{37}${K} from an
  on-line {I}sotope {S}eparator}.
\newblock \emph{\bibinfo{journal}{Phys. Rev. Lett.}}
  \textbf{\bibinfo{volume}{79}}, \bibinfo{pages}{375--378}
  (\bibinfo{year}{1997}).
\newblock \urlprefix\url{https://link.aps.org/doi/10.1103/PhysRevLett.79.375}.

\bibitem{Martensson_Pendrill_1990}
\bibinfo{author}{Martensson-Pendrill, A.~M.}, \bibinfo{author}{Pendrill, L.},
  \bibinfo{author}{Salomonson, A.}, \bibinfo{author}{Ynnerman, A.} \&
  \bibinfo{author}{Warston, H.}
\newblock \bibinfo{title}{Reanalysis of the isotope shift and nuclear charge
  radii in radioactive potassium isotopes}.
\newblock \emph{\bibinfo{journal}{J. Phys. B: At. Mol. Opt. Phys}}
  \textbf{\bibinfo{volume}{23}}, \bibinfo{pages}{1749--1761}
  (\bibinfo{year}{1990}).
\newblock
  \urlprefix\url{https://iopscience.iop.org/article/10.1088/0953-4075/23/11/012/meta}.

\bibitem{Sahoo2020}
\bibinfo{author}{Sahoo, B.~K.} \emph{et~al.}
\newblock \bibinfo{title}{Analytic response relativistic coupled-cluster
  theory: the first application to indium isotope shifts}.
\newblock \emph{\bibinfo{journal}{New J. Phys.}} \textbf{\bibinfo{volume}{22}},
  \bibinfo{pages}{012001} (\bibinfo{year}{2020}).
\newblock \urlprefix\url{https://doi.org/10.1088%2F1367-2630%2Fab66dd}.

\bibitem{scrit}
\bibinfo{author}{Ohnishi, T.} \emph{et~al.}
\newblock \bibinfo{title}{The {SCRIT} electron scattering facility project at
  {RIKEN} {RI} beam factory}.
\newblock \emph{\bibinfo{journal}{Physica Scripta}}
  \textbf{\bibinfo{volume}{T166}}, \bibinfo{pages}{014071}
  (\bibinfo{year}{2015}).
\newblock
  \urlprefix\url{https://doi.org/10.1088%2F0031-8949%2F2015%2Ft166%2F014071}.

\bibitem{perrot2006}
\bibinfo{author}{Perrot, F.} \emph{et~al.}
\newblock \bibinfo{title}{\ensuremath{\beta}-decay studies of neutron-rich {K}
  isotopes}.
\newblock \emph{\bibinfo{journal}{Phys. Rev. C}} \textbf{\bibinfo{volume}{74}},
  \bibinfo{pages}{014313} (\bibinfo{year}{2006}).
\newblock \urlprefix\url{https://link.aps.org/doi/10.1103/PhysRevC.74.014313}.

\bibitem{Hey2016}
\bibinfo{author}{Heylen, H.} \emph{et~al.}
\newblock \bibinfo{title}{Changes in nuclear structure along the {M}n isotopic
  chain studied via charge radii}.
\newblock \emph{\bibinfo{journal}{Phys. Rev. C}} \textbf{\bibinfo{volume}{94}},
  \bibinfo{pages}{054321} (\bibinfo{year}{2016}).
\newblock \urlprefix\url{https://link.aps.org/doi/10.1103/PhysRevC.94.054321}.

\bibitem{AE_sat}
\bibinfo{author}{Ekstr\"om, A.} \emph{et~al.}
\newblock \bibinfo{title}{Accurate nuclear radii and binding energies from a
  chiral interaction}.
\newblock \emph{\bibinfo{journal}{Phys. Rev. C}} \textbf{\bibinfo{volume}{91}},
  \bibinfo{pages}{051301} (\bibinfo{year}{2015}).
\newblock \urlprefix\url{https://link.aps.org/doi/10.1103/PhysRevC.91.051301}.

\bibitem{Fayan2017}
\bibinfo{author}{Reinhard, P.-G.} \& \bibinfo{author}{Nazarewicz, W.}
\newblock \bibinfo{title}{Toward a global description of nuclear charge radii:
  {E}xploring the {F}ayans energy density functional}.
\newblock \emph{\bibinfo{journal}{Phys. Rev. C}} \textbf{\bibinfo{volume}{95}},
  \bibinfo{pages}{064328} (\bibinfo{year}{2017}).
\newblock \urlprefix\url{https://link.aps.org/doi/10.1103/PhysRevC.95.064328}.

\bibitem{Hamm18a}
\bibinfo{author}{Hammen, M.} \emph{et~al.}
\newblock \bibinfo{title}{From calcium to cadmium: Testing the pairing
  functional through charge radii measurements of
  $^{100\text{\ensuremath{-}}130}\mathrm{Cd}$}.
\newblock \emph{\bibinfo{journal}{Phys. Rev. Lett.}}
  \textbf{\bibinfo{volume}{121}}, \bibinfo{pages}{102501}
  (\bibinfo{year}{2018}).
\newblock
  \urlprefix\url{https://link.aps.org/doi/10.1103/PhysRevLett.121.102501}.

\bibitem{GINS2018286}
\bibinfo{author}{Gins, W.} \emph{et~al.}
\newblock \bibinfo{title}{Analysis of counting data: Development of the satlas
  python package}.
\newblock \emph{\bibinfo{journal}{Comput. Phys. Commun.}}
  \textbf{\bibinfo{volume}{222}}, \bibinfo{pages}{286 -- 294}
  (\bibinfo{year}{2018}).
\newblock
  \urlprefix\url{http://www.sciencedirect.com/science/article/pii/S0010465517302990}.

\bibitem{Audi}
\bibinfo{author}{Wang, M.} \emph{et~al.}
\newblock \bibinfo{title}{The {AME}2016 atomic mass evaluation ({II}). tables,
  graphs and references}.
\newblock \emph{\bibinfo{journal}{Chin. Phys. C}}
  \textbf{\bibinfo{volume}{41}}, \bibinfo{pages}{030003}
  (\bibinfo{year}{2017}).
\newblock
  \urlprefix\url{https://iopscience.iop.org/article/10.1088/1674-1137/41/3/030003/meta}.

\bibitem{fricke2004nuclear}
\bibinfo{author}{Fricke, G.} \& \bibinfo{author}{Heilig, K.}
\newblock \bibinfo{title}{Nuclear charge radii in \lq\lq elementary particles,
  nuclei and atoms\rq\rq}.
\newblock
  \urlprefix\url{https://materials.springer.com/lb/docs/sm_lbs_978-3-540-45555-4_21}.
\newblock \bibinfo{note}{Springer-Verlag Berlin Heidelberg}.

\bibitem{Bendali_1981}
\bibinfo{author}{Bendali, N.}, \bibinfo{author}{Duong, H.~T.} \&
  \bibinfo{author}{Vialle, J.~L.}
\newblock \bibinfo{title}{High-resolution laser spectroscopy on the {D1} and
  {D2} lines of $^{39,40,41}\mathrm{K}$ using {RF} modulated laser light}.
\newblock \emph{\bibinfo{journal}{J. Phys. B: At. Mol. Opt. Phys.}}
  \textbf{\bibinfo{volume}{14}}, \bibinfo{pages}{4231--4240}
  (\bibinfo{year}{1981}).
\newblock
  \urlprefix\url{https://iopscience.iop.org/article/10.1088/0022-3700/14/22/009/meta}.

\bibitem{Comb}
\bibinfo{author}{Falke, S.}, \bibinfo{author}{Tiemann, E.},
  \bibinfo{author}{Lisdat, C.}, \bibinfo{author}{Schnatz, H.} \&
  \bibinfo{author}{Grosche, G.}
\newblock \bibinfo{title}{Transition frequencies of the ${D}$ lines of
  $^{39}\mathrm{K}$, $^{40}\mathrm{K}$, and $^{41}\mathrm{K}$ measured with a
  femtosecond laser frequency comb}.
\newblock \emph{\bibinfo{journal}{Phys. Rev. A}} \textbf{\bibinfo{volume}{74}},
  \bibinfo{pages}{032503} (\bibinfo{year}{2006}).
\newblock \urlprefix\url{https://link.aps.org/doi/10.1103/PhysRevA.74.032503}.

\bibitem{Linear_error}
\bibinfo{author}{Petersen, P.} \emph{et~al.}
\newblock \bibinfo{title}{Models for combining random and systematic errors.
  assumptions and consequences for different models.}
\newblock \emph{\bibinfo{journal}{Clin. Chem. Lab. Med.}}
  \textbf{\bibinfo{volume}{39}}, \bibinfo{pages}{589--595}
  (\bibinfo{year}{2001}).

\bibitem{tichai2019}
\bibinfo{author}{Tichai, A.}, \bibinfo{author}{M\"uller, J.},
  \bibinfo{author}{Vobig, K.} \& \bibinfo{author}{Roth, R.}
\newblock \bibinfo{title}{Natural orbitals for ab initio no-core shell model
  calculations}.
\newblock \emph{\bibinfo{journal}{Phys. Rev. C}} \textbf{\bibinfo{volume}{99}},
  \bibinfo{pages}{034321} (\bibinfo{year}{2019}).
\newblock \urlprefix\url{https://link.aps.org/doi/10.1103/PhysRevC.99.034321}.

\bibitem{bartlett2007}
\bibinfo{author}{Bartlett, R.~J.} \& \bibinfo{author}{Musia\l{}, M.}
\newblock \bibinfo{title}{Coupled-cluster theory in quantum chemistry}.
\newblock \emph{\bibinfo{journal}{Rev. Mod. Phys.}}
  \textbf{\bibinfo{volume}{79}}, \bibinfo{pages}{291--352}
  (\bibinfo{year}{2007}).
\newblock \urlprefix\url{http://link.aps.org/doi/10.1103/RevModPhys.79.291}.

\bibitem{AE_delta}
\bibinfo{author}{Ekstr\"om, A.}, \bibinfo{author}{Hagen, G.},
  \bibinfo{author}{Morris, T.~D.}, \bibinfo{author}{Papenbrock, T.} \&
  \bibinfo{author}{Schwartz, P.~D.}
\newblock \bibinfo{title}{$\mathrm{\ensuremath{\Delta}}$ isobars and nuclear
  saturation}.
\newblock \emph{\bibinfo{journal}{Phys. Rev. C}} \textbf{\bibinfo{volume}{97}},
  \bibinfo{pages}{024332} (\bibinfo{year}{2018}).
\newblock \urlprefix\url{https://link.aps.org/doi/10.1103/PhysRevC.97.024332}.

\bibitem{Siemens17}
\bibinfo{author}{Siemens, D.} \emph{et~al.}
\newblock \bibinfo{title}{Reconciling threshold and subthreshold expansions for
  pion–nucleon scattering}.
\newblock \emph{\bibinfo{journal}{Phys. Lett. B}}
  \textbf{\bibinfo{volume}{770}}, \bibinfo{pages}{27 -- 34}
  (\bibinfo{year}{2017}).
\newblock
  \urlprefix\url{http://www.sciencedirect.com/science/article/pii/S0370269317303131}.

\bibitem{Hebeler2011}
\bibinfo{author}{Hebeler, K.}, \bibinfo{author}{Bogner, S.~K.},
  \bibinfo{author}{Furnstahl, R.~J.}, \bibinfo{author}{Nogga, A.} \&
  \bibinfo{author}{Schwenk, A.}
\newblock \bibinfo{title}{Improved nuclear matter calculations from chiral
  low-momentum interactions}.
\newblock \emph{\bibinfo{journal}{Phys. Rev. C}} \textbf{\bibinfo{volume}{83}},
  \bibinfo{pages}{031301} (\bibinfo{year}{2011}).
\newblock \urlprefix\url{https://link.aps.org/doi/10.1103/PhysRevC.83.031301}.

\bibitem{Fayans1998}
\bibinfo{author}{Fayans, S.~A.}
\newblock \bibinfo{title}{Towards a universal nuclear density functional}.
\newblock \emph{\bibinfo{journal}{JETP Lett.}} \textbf{\bibinfo{volume}{68}},
  \bibinfo{pages}{169} (\bibinfo{year}{1998}).
\newblock \urlprefix\url{http://dx.doi.org/10.1134/1.567841}.

\bibitem{Minamisono16}
\bibinfo{author}{Minamisono, K.} \emph{et~al.}
\newblock \bibinfo{title}{Charge radii of neutron deficient
  $^{52,53}\mathrm{Fe}$ produced by projectile fragmentation}.
\newblock \emph{\bibinfo{journal}{Phys. Rev. Lett.}}
  \textbf{\bibinfo{volume}{117}}, \bibinfo{pages}{252501}
  (\bibinfo{year}{2016}).
\newblock
  \urlprefix\url{http://link.aps.org/doi/10.1103/PhysRevLett.117.252501}.

\bibitem{STOITSOV20131592}
\bibinfo{author}{Stoitsov, M.} \emph{et~al.}
\newblock \bibinfo{title}{Axially deformed solution of the
  skyrme-hartree-fock-bogoliubov equations using the transformed harmonic
  oscillator basis (ii) hfbtho v2.00d: A new version of the program}.
\newblock \emph{\bibinfo{journal}{Comput. Phys. Comm.}}
  \textbf{\bibinfo{volume}{184}}, \bibinfo{pages}{1592 -- 1604}
  (\bibinfo{year}{2013}).
\newblock
  \urlprefix\url{http://www.sciencedirect.com/science/article/pii/S0010465513000301}.

\end{thebibliography}

\vspace{10mm}

\centerline{\textbf{METHODS}}
\vspace{3mm}

\textbf{The CRIS technique:}  The schematic layout of the CRIS setup is
presented in Fig.~\ref{CRIS}. The mass selected ions were cooled and bunched in the ISCOOL device which operated at 100 Hz, the duty cycle of the CRIS experiment. The ion bunches typically have a \mbox{6 \,$\mu s$} temporal width corresponding to a spatial length of around 1\,m length. First, the bunched ion beam was neutralized in the CRIS beamline through collisions with potassium atoms in the charge-exchange cell (CEC). The remaining ions were deflected just after the CEC with an electrostatic deflector plate. In addition, atoms that were produced in highly
excited states through the charge exchange process were field ionized and deflected out of the beam. The beam of neutral atoms passed through the differential pumping region and arrived in the \mbox{1.2-m} interaction
region (IR) maintained at a pressure of $10^{-10}$~mbar. Here the atom
bunch was collinearly overlapped with 3 laser pulses, which were used to step-wise
excite and ionize the potassium atoms. A detailed study of this particular resonance ionization scheme can be found in Ref.~\cite{AGIPRC2019}. In the IR,
ions can also be produced in non-resonant processes, introducing higher
background rates~\cite{KOSZORUS2019}. Normally, the ions created in
the IR are guided towards an ion detector (MagneToF). This technique was used for the measurement of $^{47-51}$K. The $^{51}$K isotope was produced at a rate of less than 2000 particles per second and its hfs spectrum was measured in less than 2 hours. The study of $^{52}$K, however, required a still more selective detection method. This isotope, produced at about 360 particles per second rate, is an isobar of the most abundant stable chromium isotope, which is the main contaminating species in the $A=52$ beam, with an intensity of 6$\times 10^{6}$ particles per second.
In order to avoid the detection of the non-resonantly ionized $^{52}$Cr, the CRIS setup was equipped with a decay detection station, placed behind the end flange of the beamline. The MagneToF detector was removed from the path of the ion beam, and the ionized bunches were implanted into a thin aluminium window with 1~mm thickness allowing the transmission of $\beta$
particles with energies larger than 0.6~MeV. The decay station
behind this window consisted of a thin and thick scintillator detector (A and B in Fig.~\ref{CRIS}) for a coincidence detection of $\beta$-particles. The dimensions of the detectors were $1\,\text{mm}\times 6\,\text{cm}\times 6\,\text{cm}$ and $6\,\text{cm}\times 6\,\text{cm}\times 6\,\text{cm}$. The $\beta$ counts detected in coincidence were recorded by the data acquisition system (DAQ) together with the laser frequency detuning. The DAQ recorded the number of events in the detectors with a timestamp. The timestamp of the proton bunches impinging into the target of ISOLDE was also recorded and used to define the time gates in the data analysis.

\vspace{3mm}
\textbf{Laser system:} A three-step resonance ionization scheme was
used in this experiment. The laser light for the first excitation step was
produced by a continuous wave (cw) titanium-sapphire (Ti:Sa) laser
\mbox{(M-Squared SolsTiS)} pumped by an 18-W laser at 532\,nm
\mbox{(Lighthouse Photonics)}. In order to avoid optical pumping to dark
states due to long interaction times, this cw light was \lq\lq chopped'' into 50-ns pulses at a
repetition rate of \mbox{100~Hz} by using a Pockels cell \cite{RubenPRL}.
The wavelength of this narrowband laser was tuned to probe the hfs of the \mbox{4s
  $^{2}$S$_{1/2}$ $\rightarrow$ 4p $^{2}$P$_{1/2}$} transition at
769~nm. Atoms in the excited \mbox{4p $^{2}$P$_{1/2}$} state were
subsequently further excited to the \mbox{6s $^{2}$S$_{1/2}$} atomic state by a pulsed dye laser \mbox{(Spectron PDL SL4000)} with a spectral bandwidth of
10~GHz. This dye laser was pumped by a 532 nm Nd:YAG laser \mbox{(Litron TRLi 250-100)} at a 100-Hz repetition rate. The fundamental output of the same Nd:YAG laser (1064~nm) was
used for the final non-resonant ionization. The arrival of ion
bunches and laser pulses in the interaction region were synchronized and
controlled using a multi-channel pulse
generator (Quantum Composers 9520 Series).
%The timing of each laser pulse could be adjusted independently, allowing the systematic investigation of the effect of different laser pulse sequences on the spectral line shape and extracted observables from experimental hfs spectra \cite{AGIPRC2019}.
%vspace{5mm}\\

\vspace{3mm} \textbf{Charge radii extraction:}\label{rms_calc} The
perturbation of the atomic states caused by the different nuclear
charge distribution in isotopes leads to small differences in the atomic
transition frequency, $\delta\nu^{AA'}$, between centroids
($\nu^{A}$,$\nu^{A'}$) of the hfs of two isotopes with mass number $A$
and $A'$. The isotope shifts of $^{48-52}$K were extracted from the hfs
spectra of $^{47-52}$K analyzed using the SATLAS \cite{GINS2018286}
Python package, as displayed in third column of Table \ref{table_evaluation}, along with all available results from literature. More details on the analysis process
can be found in Ref.\cite{AGIPRC2019}. The changes in the nuclear
mean-square charge radii of $^{36-52}$K can then be extracted from the isotope shifts using:
\begin{equation}
\delta\nu^{AA'} = K_{\rm MS} \frac{m_{A}-m_{A'}}{m_{A'}(m_{A}+m_{e})} + F \delta {\langle {r}^{2} \rangle}^{AA'},
\label{IS}
\end{equation}
where $K_{\rm MS}$ and $F$ are the atomic mass shift and field shift, $m$
stands for the nuclear mass of isotopes $A$, $A'$, and an electron. The nuclear mass was obtained by subtracting the mass of the electrons from the experimentally measured atomic mass reported in Ref.\cite{Audi}. The atomic
constants, $K_{\rm MS}$ and $F$, were calculated using the atomic
ARRCC method as described below. The root-mean-square
charge radii of these isotopes are:
\begin{equation}
R = \sqrt{\delta \langle r^{2} \rangle + R_{39}^{2}},
\label{rad}
\end{equation}
where $R_{39}$ is the charge radius of $^{39}$K taken from Ref. \cite{fricke2004nuclear}.

\begin{table*}[t!]
\caption{Experimental re-evaluated isotope shifts, changes in mean-square charge radii and the absolute charge radii of $^{36-52}$K. Experimental and theoretical systematic errors are reported in curly and square brackets, respectively. The statistical uncertainties are presented in brackets. The reference value for the charge radius of $^{39}$K is taken from Ref. \cite{fricke2004nuclear}, where it is given without uncertainties.}\label{table_evaluation}
\vspace{2mm}
\renewcommand*{\arraystretch}{1.2}
\begin{tabular}{p{0.8cm} p{2.5cm} p{2.5cm}p{1.8cm}p{2.2cm}|p{2.2cm}p{2.2cm}p{2.8cm}}
 A& $\delta \nu^{39,A}$ (MHz) &  $\delta \nu^{47,A}$ (MHz)&  Reference &$\delta \nu^{39,A}_{\rm re}$ (MHz) &  $\overline{\delta \nu}^{39,A}$ & $\delta\langle r^{2}\rangle^{39,A}$ (fm$^2$) & $R_{\rm ch}$ (fm)\\
\hline
36        & -403(5)\{4\}           &                   &  Ref. \cite{Rossi}          &            & -403(9)    &-0.20(8)[4] & 3.405(12)[6] \\
\hline
37        & -264(3)\{3\}           &                   & Ref. \cite{Rossi}           &            & -264(6)    &-0.11(6)[3]& 3.419(8)[4] \\
%37& -265(4)           &    & Ref. \cite{Behr}         & & \\
\hline
38        & -127.0(53)             &                   & Ref. \cite{TOUCHARD1982169} &            &   \multirow{3}{*}{-126.1(19)} & \multirow{3}{*}{-0.075(18)[14]}  & \multirow{3}{*}{3.4241(26)[20]} \\
38        &                        &  -985.9(4)\{34\}  & Ref. \cite{KREIM201497}     &  -127.5(39)   &            &  \\
38        &                        &  -983.8(4)\{18\}  & Ref. \cite{AGIPRC2019}      &  -125.4(24)&            &  \\
\hline
39        &                        &  -862.5(9)\{30\}  & Ref. \cite{KREIM201497}     &            &      \multirow{2}{*}{0}     & \multirow{2}{*}{0} & \multirow{2}{*}{3.435}\\
39        &                        &  -858.4(6)\{5\}   & Ref. \cite{AGIPRC2019}      &            &            &    \\
\hline
40        & 125.58(26)             &                   & Ref. \cite{Bendali_1981}    &            & \multirow{2}{*}{125.63(9)} & \multirow{2}{*}{0.025(1)[13]}&\multirow{2}{*}{3.4386(1)[19]}   \\
40        & 125.64(10)             &                   &Ref. \cite{Comb}             &            &            &   \\
\hline
41         & 235.25(75)            &                   & Ref. \cite{TOUCHARD1982169} &            &\multirow{3}{*}{235.47(9)}   & \multirow{3}{*}{0.135(1)[26]}&\multirow{3}{*}{3.4546(1)[37]}    \\
41         & 235.27(33)            &                   &Ref. \cite{Bendali_1981}     &            &            &     \\
41         & 235.49(9)             &                   &Ref. \cite{Comb}             &            &            &  \\
\hline
42         & 351.7(19)             &                   & Ref. \cite{TOUCHARD1982169} &            & \multirow{3}{*}{352.4(10)}  &\multirow{3}{*}{0.128(10)[38]}&\multirow{3}{*}{3.4536(15)[55]}  \\
42         &                       &  -506.7(7)\{17\}  & Ref. \cite{KREIM201497}     &   351.7(26)&            &  \\
42         &                       &   -505.5(6)\{3\}  & Ref. \cite{AGIPRC2019}      &   352.9(13)&            &      \\
\hline
43          &459.0(12)             &                   &  Ref. \cite{TOUCHARD1982169}&            & 459.0(12)  & 0.165(11)[49]& 3.4590(16)[71] \\
\hline
44          & 564.3(14)            &                   & Ref. \cite{TOUCHARD1982169} &            & \multirow{3}{*}{565.1(8)}   & \multirow{3}{*}{0.163(7)[60]} &\multirow{3}{*}{3.4586(11)[87]} \\
44          &                      & -292.1(5)\{10\}   & Ref. \cite{KREIM201497}     &   566.3(17)&            &  \\
44          &                      & -293.19(56)\{23\} & Ref. \cite{AGIPRC2019}      &   565.2(12)&            &  \\
\hline
45          & 661.7(16)            &                   &Ref. \cite{TOUCHARD1982169}  &            & 661.7(16)  &0.203(15)[70] & 3.4644(22)[102]  \\
\hline
46          & 762.8(15)            &                   &Ref. \cite{TOUCHARD1982169}  &            &  \multirow{3}{*}{764.1(14)}       &  \multirow{3}{*}{0.150(13)[80]}  &\multirow{3}{*}{3.4568(19)[116]} \\
46          &                      &-91.6(5)\{3\}      &Ref. \cite{KREIM201497}      &   766.8(12)&            &  \\
46          &                      &-95.81(55)\{6\}    &Ref. \cite{AGIPRC2019}       &   762.6(11)&            &  \\
\hline

47          &    857.5(17)         &                   &  Ref. \cite{TOUCHARD1982169}&            &     $\bm{858.4(9)}$        & 0.133(8)[90]& 3.4543(12)[130] \\
\hline
48          &                      & 67.9(4)\{3\}      & Ref. \cite{KREIM201497}     &   926.3(11)&   \multirow{2}{*}{926.4(9)}    & \multirow{2}{*}{0.328(8)[99]}  &\multirow{2}{*}{3.4825(12)[142]}  \\
48          &                      & 68.2(11)          &  This work                  &   926.6(15)&            &  \\
\hline
49          &                      & 135.3(5)\{6\}     &Ref.  \cite{KREIM201497}     &   993.7(14)&\multirow{2}{*}{994.2(11)}  &  \multirow{2}{*}{0.491(10)[108]}   &\multirow{2}{*}{3.5057(15)[154]}  \\
49          &                      & 136.4(13)         & This work                   &   994.8(16)&                              &  \\
\hline
50          &                      & 206.5(9)\{9\}     &Ref.  \cite{KREIM201497}     &   1064.9(20)&\multirow{2}{*}{1065.3(10)}   &\multirow{2}{*}{0.592(9)[116]}   &\multirow{2}{*}{3.5201(13)[165]} \\
50          &                      & 207.1(8)          &This work                    &   1065.5(12)&                              & \\
\hline
51         &                      & 273.2(14)\{11\}    &Ref. \cite{KREIM201497}      &   1131.6(27)& \multirow{2}{*}{1130.6(14)}  &\multirow{2}{*}{0.716(13)[124]}  &\multirow{2}{*}{3.5377(18)[176]} \\
51         &                      & 271.9(13)          & This work                   &   1130.3(16)&             & \\
\hline
52         &                      & 340(5)             &This work                    &   1198(5)   &  1198(5)   &0.79(5)[132]&  3.549(7)[19]\\
\hline
\end{tabular}
\end{table*}

\vspace{3mm}
\textbf{Evaluation of the isotope shifts and charge radii:} The isotope shifts of potassium isotopes were measured using several different techniques over many years, ranging from magneto-optical trap experiments \cite{Behr} to laser spectroscopy of thermal \cite{Comb, TOUCHARD1982169,Bendali_1981} and accelerated beams \cite{Rossi,KREIM201497,AGIPRC2019}, relying on photon and ion detection. The available results in Refs. \cite{Behr,Comb, TOUCHARD1982169, Bendali_1981, Rossi} are referenced to the stable $^{39}$K isotope, and are presented in the second column of Table \ref{table_evaluation}. The isotope shifts in Refs. \cite{KREIM201497, AGIPRC2019} and this work, shown in the third column of Table \ref{table_evaluation}, were extracted with respect to $^{47}$K. The systematic error from the experiments is given in curly brackets. Note that the systematic uncertainties in collinear laser spectroscopy experiments are mostly related to the inaccuracy of the acceleration voltage. In this work, the systematic uncertainty was negligible by using the laser scanning approach \cite{AGIPRC2019} and a well-calibrated high-precision voltage divider (with a relative uncertainty of $5 \times 10^{-5}$) from PTB. % compared with the divider used in Ref.\cite{AGIPRC2019}.
In order to compile a consistent data set with reliable evaluation of uncertainties, the following steps were taken:
\begin{itemize}
\item [1)]
The isotope shifts obtained with respect to $^{47}$K were recalculated relative to $^{39}$K, in order to link all data to the same reference. For this, the weighted average of all available $\delta\nu^{47,39}$ isotope shifts from Refs. \cite{KREIM201497, AGIPRC2019} is used as a reference. These re-referenced values are listed in the fifth column of Table \ref{table_evaluation} and their uncertainty is increased due to the additional error associated with $\overline{\delta \nu}^{39,47}$ (bold value in column six). Note that the systematic errors are always taken into account using the linear model \cite{Linear_error}, $\sigma= \sigma_{\rm sys}+\sigma_{\rm stat}$.
\vspace{-1mm}
\item [2)]
Next, the final isotope shift of each potassium isotope, $\overline{\delta \nu}^{39,A}$ (shown in the sixth column of Table \ref{table_evaluation}), was calculated as the weighted average ($\hat{x}$) of the available results ($x_{i}$) using:
%Eq.\ref{average} only use EQ.4 when already defined
\begin{equation}
\hat{x} = \frac {\Sigma_{i=1}^{n}(x_{i}\sigma_{i}^{-2})} {\Sigma_{i=1}^{n}\sigma_{i}^{-2}},
\label{average}
\end{equation}
where $\sigma_{i}$ is the total uncertainty of $i^{\rm th}$ measurement.
The error of the weighted mean was obtained using:
%Eq. \ref{error}
\begin{equation}
\sigma_{\hat{x}}= \sqrt{\frac{1} {\Sigma_{i=1}^{n} \sigma_{i}^{-2}}},
\label{error}
\end{equation}
accounting for possible over-, or under-dispersion using: % Eq. \ref{correction}.
\begin{equation}
\hat{\sigma}_{\overline{x}}^{2} = \sigma_{\overline{x}}^{2}\chi^{2},
\label{correction}
\end{equation}
where $\chi^{2}$ is the reduced chi-squared.
\item [3)]
These evaluated isotope shifts were used to extract the changes in mean-square charge radii (column 7 of Table \ref{table_evaluation}) using the new theoretical values for the atomic field and mass shifts, obtained in this work. The absolute charge radii of all potassium isotopes (last column of Table \ref{table_evaluation}) were then calculated by using the above mentioned \mbox{Eq.\ref{rad}}, relative to the absolute radius of the stable $^{39}$K \cite{fricke2004nuclear}.
\end{itemize}

\vspace{3mm}
\textbf{Atomic coupled-cluster calculations:} The wave function of an atomic state with a closed-shell and a valence orbital electronic configuration can be expressed using the CC theory {\it ansatz} as
\begin{eqnarray}
|\Psi_v \rangle &=& e^{\tilde S} |\Phi_v \rangle= e^T \{1+S_v\} |\Phi_v \rangle,
\end{eqnarray}
where $|\Phi_v \rangle$ is the mean-field wave function and ${\tilde
  S}$ is the CC excitation operator. We further divide as ${\tilde
  S}=T+S_v$ to distinguish electron correlations without involving the
valence electron ($T$) and involving the valence electron ($S_v$). In
the analytic response procedure, the first-order energy of the atomic
state is obtained by solving the equation
\begin{eqnarray}
(H_{\rm at}-E_v^{(0)} ) |\Psi_v^{(0)} \rangle = (E_v^{(1)}- H_{\rm int} ) |\Psi_v^{(1)}\rangle ,
\end{eqnarray}
where $H_{\rm at}$ is the atomic Hamiltonian, $H_{\rm int}$ is the interaction
Hamiltonian, $|\Psi_v^{(0)} \rangle$ is the unperturbed wave function
with energy $E_v^{(0)}$, and $|\Psi_v^{(1)} \rangle$ is the
first-order perturbed wave function with the first-order energy
$E_v^{(1)}$. Here, $H_{\rm at}$ involves the Dirac terms, the nuclear
potential, the lower-order quantum electrodynamics corrections, and
the electron-electron interactions due to the longitudinal and
transverse photon exchanges, while $H_{\rm int}$ is either the FS operator
due to the Fermi nuclear charge distribution in the evaluation of $F$
or the relativistic form of the SMS operator for the determination of
$K_{\rm SMS}$. In the ARRCC theory, the unperturbed and the perturbed wave
functions are obtained by expanding
\begin{eqnarray}
T \simeq T^{(0)} + \lambda T^{(1)} \ \ \ \ \text{and} \ \ \ \ S_v \simeq S_v^{(0)} + \lambda S_v^{(1)}
\end{eqnarray}
with $\lambda$ representing the perturbation parameter. After
solving the amplitudes of both the unperturbed and perturbed CC
operators, as described in Ref. \cite{Sahoo2020}, we evaluate the
first-order energy as
\begin{eqnarray}
E_v^{(1)} &=&  \langle \Phi_v | ( H_{\rm at} e^{T^{(0)}} )_c  \{ S_v^{(1)} +  T^{(1)} \} | \Phi_v \rangle \nonumber \\ && + \langle \Phi_v |( H_{\rm int} e^{T^{(0)}} )_c \{ 1+ S_v^{(0)} \}  | \Phi_v \rangle ,
\end{eqnarray}
in which the subscript $c$ means the connected terms. We have
considered all possible single, double and triple electronic
excitation configurations in our ARRCC method for performing the
atomic calculations.

\begin{figure}[!t]%fig1
\begin{center}
\label{fig:K-isotopes-isotopeshift}\includegraphics[width=.49\textwidth]{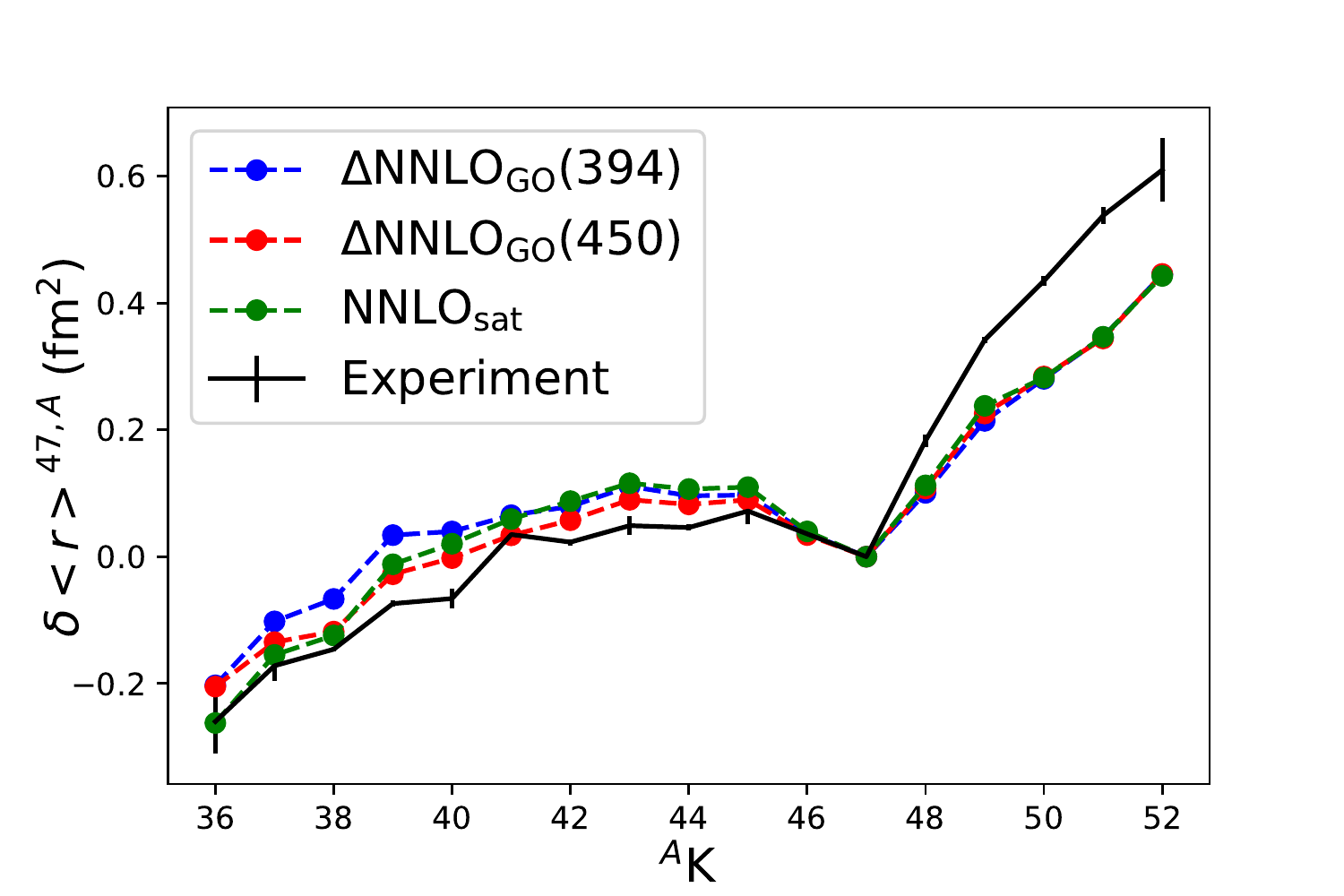}\\
\label{fig:potassium_isotope_shift}\includegraphics[width=.50\textwidth]{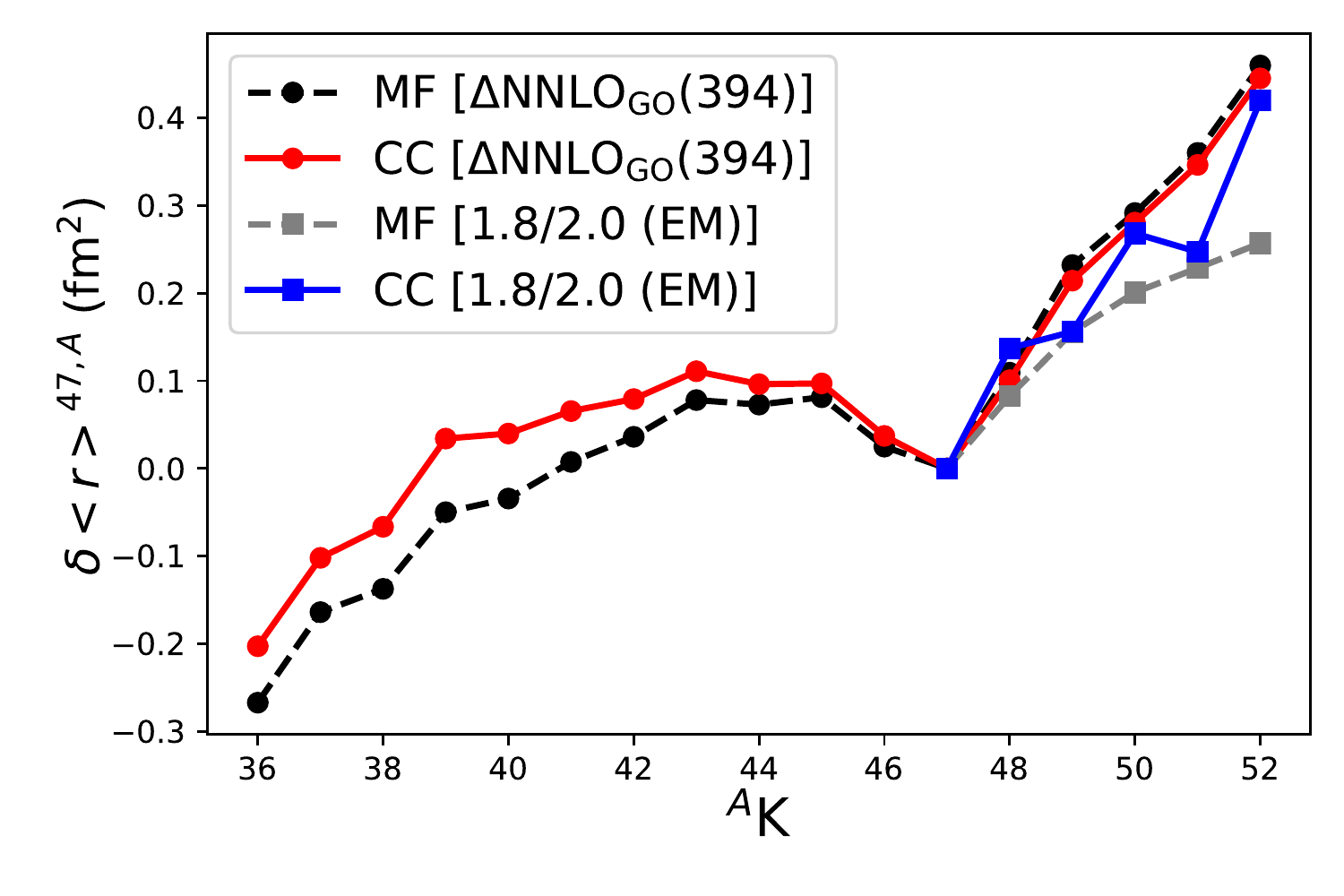}
\vspace{-3mm}
\caption{\footnotesize{Changes in mean-square charge radii of potassium isotopes calculated based on newly developed interactions and NNLO$_{\rm sat}$ are compared with (a) experimental data and (b) that from deformed mean-field (MF) calculations with CCSD computations for two different interactions.}}
\label{radii}
\end{center}
\end{figure}
\vspace{3mm}
\textbf{Nuclear coupled cluster calculations:} The nuclear CC calculations start from the intrinsic Hamiltonian including two- and three-nucleon forces,
\begin{equation}
  \label{intham}
  \hat{H} = \sum_{i<j}\left({({\bf p}_i-{\bf p}_j)^2\over 2mA} + \hat{V}
    _{NN}^{(i,j)}\right) + \sum_{ i<j<k}\hat{V}_{\rm 3N}^{(i,j,k)}.
\end{equation}
The nuclear CC wave function is written as
$\vert \Psi\rangle = e^{T}\vert \Phi_0\rangle$, where the cluster
operator $T$ is a linear combination of $n$-particle-$n$-hole
excitations truncated at the two-particle-two-hole level, commonly
known as the CC with singles-and-double excitations
(CCSD). The reference state $\vert \Phi_0 \rangle$ is restricted to
axial symmetry and is constructed in the following way. We start by
solving the self-consistent Hartree-Fock equations by assuming that
the most energetically favorable configuration is obtained by filling
the states with the lowest angular momentum projection along the
$z$-axis (prolate deformation). Subsequently we construct a natural
orbital basis by diagonalizing the density matrix obtained from
second-order perturbation theory~\cite{tichai2019}. Finally, we
normal-order the Hamiltonian (\ref{intham}) with respect to the
natural orbital mean-field solution keeping up to two-body
normal-ordered contributions. The model-space used in our calculations
is given by 13 major harmonic oscillator shells ($N_{\rm max} = 12$)
with the oscillator frequency $\hbar\Omega = 16$~MeV. The three-body
interaction has the additional cutoff on allowed three-particle
configurations $E_{3\text{max}}=N_1+N_2+N_3 \leq 16$, with
$N_i = 2n_i +l_i$. This model-space is sufficient to converge the radii
of all the potassium isotopes considered in this work to within
$\sim$1\% . In this work we calculate the expectation value of the
squared intrinsic point proton radius, i.e.
$\langle O\rangle  = \langle {1/Z} \sum_{i<j} ( {\bf r}_i - {\bf r}_j)^2 \delta_{t_z,-1}\rangle$. The
coupled-cluster expectation value of the operator $O$ is given by,
$\langle \Phi_0 \vert \left( 1+\Lambda \right) e^{-T} O e^T \vert
\Phi_0\rangle = \langle \Phi_0 \vert \left( 1+\Lambda \right)
\overline{O} \vert \Phi_0\rangle $, here
$ \langle \Phi_0 \vert (1+\Lambda) e^{-T}$ is the left ground-state,
and $\Lambda$ is the linear combination of one-hole-one-particle and
two-hole-two-particle de-excitation operators, see
e.g. Ref.\cite{bartlett2007} for details. To obtain the charge radii we
add finite proton/neutron, the relativistic Darwin-Foldy, and
spin-orbit corrections to the point proton radii. We note that the
spin-orbit corrections are computed consistently as an expectation
value within the CC approach, see Ref. \cite{Hagen2016}.

The chiral interactions we used are NNLO$_{\rm sat}$ \cite{AE_sat}, which has
been constrained by nucleon-nucleon properties, and binding energies
and charge radii of nuclei up to oxygen. It includes terms up to
next-to-next-to leading order in the Weinberg power counting. The
newly constructed $\Delta{\rm NNLO}_{\rm GO}(450)$ interaction
includes $\Delta$ isobar degrees of freedom, exhibits a cutoff of
450~MeV, and is also limited to next-to-next-to-leading order
contributions. Its construction starts from the interaction of
Ref.~\cite{AE_delta} and its low-energy constants are constrained by the
saturation density, energy and symmetry energy of nuclear matter, by
pion-nucleon scattering \cite{Siemens17}, nucleon-nucleon scattering, and
by the $A \leq 4$ nuclei. A second interaction, $\Delta{\rm NNLO}_{\rm
GO}(394)$, was similarly constructed but with a cutoff of 394~MeV.

To look at the sensitivities in the changes in the mean-square change radii, Fig. \ref{radii}a compares
results from the newly developed interactions with NNLO$_{\rm
sat}$. While there are differences below $N=28$, all interactions
yield essentially identical results beyond $N=28$. This suggests that charge radii beyond $N = 28$ are insensitive to details of chiral interactions at next-to-next-to-leading-order.

To shed more light onto this finding, Fig. \ref{radii}b compares results from
deformed mean-field (MF) calculations with CCSD computations for two
different interactions. The 1.8/2.0(EM) interaction \cite{Hebeler2011}
contains contributions at next-to-next-to-next-to-leading order and
thereby differs from the interactions used in this work. First, for
the $\Delta{\rm NNLO}_{\rm GO}(394)$ interaction, MF and
CC yield essentially the same results for $N>28$,
though CC includes many more wave-function correlations. Second, for
the 1.8/2.0(EM) interaction, MF and CC results differ significantly by
the strong odd-even staggering, which is a correlation effect. None of
the interactions explain the dramatic increase of the charge radii
beyond $N=28$.

%\begin{figure}
%\includegraphics[width=\linewidth]{K-isotopes-isotopeshift.pdf}
%\caption{\label{fig:K-isotopes-isotopeshift}
%Charge radii of K isotopes resulted from the newly developed interactions with NNLO$_{\rm
%sat}$ and compared with experimental data.}
%\end{figure}

%\begin{figure}
%\includegraphics[width=\linewidth]{potassium_isotope_shift.pdf}
%\caption{\label{fig:potassium_isotope_shift}
%Comparison of charge radii of K isotopes resulted from the newly developed interactions with NNLO$_{\rm
%sat}$ and from deformed mean-field (MF) calculations with CCSD computations for two different interactions.}
%\end{figure}
\begin{figure}[!t]
\includegraphics[width=1\linewidth]{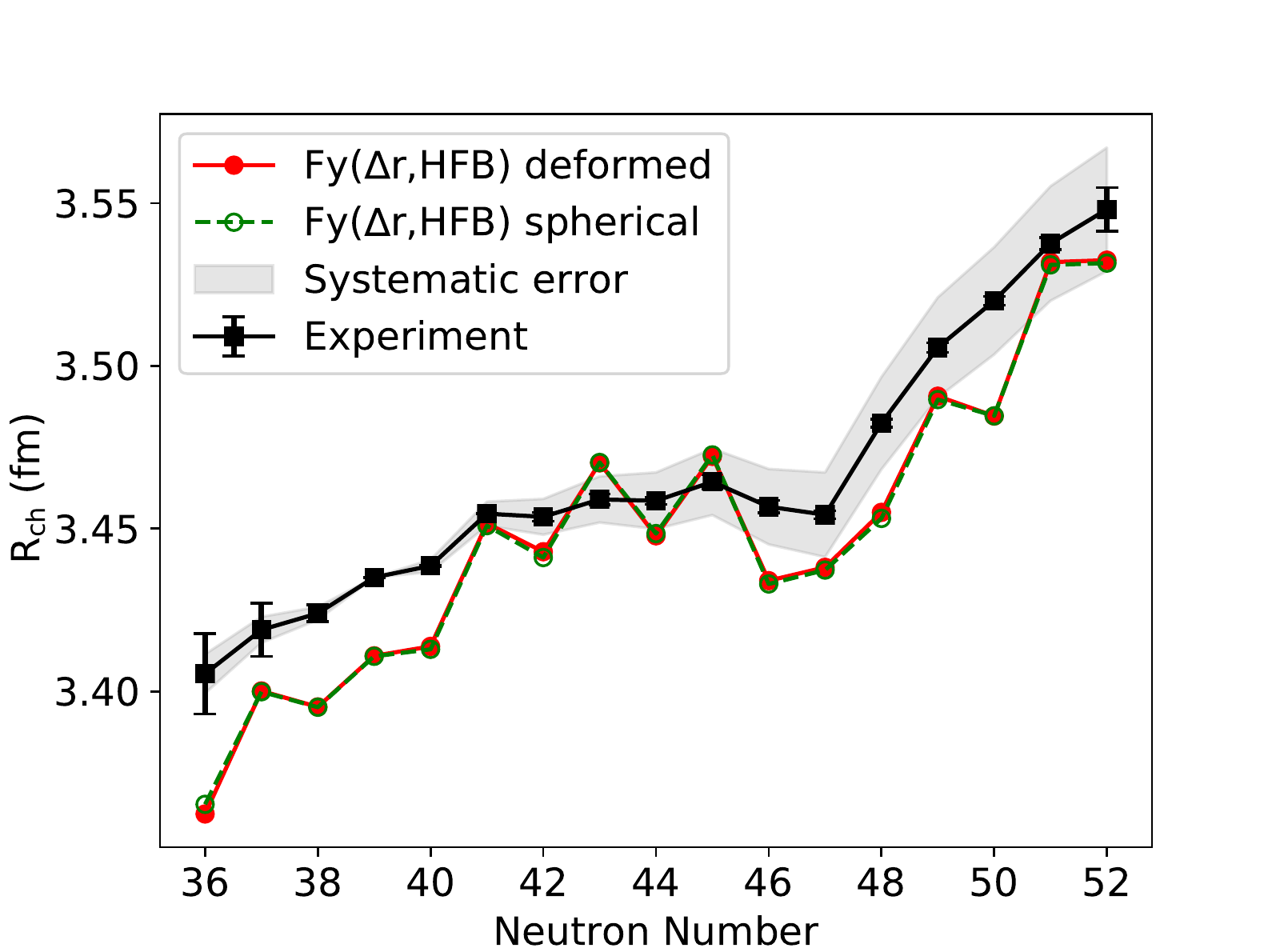}
\vspace{-3mm}
\caption{\label{fig:K-radii-FyDr-spher-def} Charge radii along the chain of potassium isotopes from spherical as well as deformed calculations with the functional Fy($\Delta r$,HFB) and compared
with experimental data.}
\vspace{-3mm}
\end{figure}

\vspace{3mm}
\textbf{DFT calculations:}
For the DFT part of this work, we use the non-relativistic Fayans
functional in the form of Ref.\cite{Fayans1998}. This functional is
distinguished from other commonly used nuclear DFT in that it has
additional gradient terms at two places, namely in the pairing
functional and in the surface energy. The gradient terms allow, among
other features, a better reproduction of the isotopic trends of charge radii
\cite{Minamisono16}. This motivated a refit of the Fayans functional
to a broad basis of nuclear ground state data with additional
information on changes in mean-square charge radii in the calcium chain
\cite{Fayan2017,GarciaRuiz2016}. We use here Fy($\Delta r$,HFB) from Refs.
\cite{Fayan2017,GarciaRuiz2016} which employed the latest data on calcium radii.
It is only with rather strong gradient terms that one is able to
reproduce the trends of radii in calcium at all, in particular its
pronounced odd-even staggering, however, with a slight tendency to
exaggerate the staggering. It was found later that Fy($\Delta r$,HFB)
performs very well in describing the trends of radii in cadmium and tin
isotopes \cite{Hamm18a,Gorges2019}. Here we test it again for the potassium
chain next to calcium.  For all practical details pertaining to our
Fayans-DFT calculations, we refer the reader to Ref. \cite{Fayan2017}.

All above mentioned calculations with Fy($\Delta r$,HFB) were done in
spherical representation. That can be questioned for nuclei far off
shell closures, the more so for odd nuclei where the odd nucleon
induces a certain quadrupole moment. The new feature in the present
calculations is that we use a code in axial representation which
allows for deformations if the systems wants any. Fig.~\ref{rmsCaK}
shows the result from the deformed code\cite{STOITSOV20131592}, adapted for the Fayans functional. Here in Fig.~\ref{fig:K-radii-FyDr-spher-def} we illustrate the effect of deformation by comparison with spherical calculations. The effects
are small, but show there is no uncertainty due to
symmetry restrictions. The lack of angular-momentum projection in DFT calculations induces a systematic error on charge radii. We have estimated it from the angular-momentum spread to be below 0.005 fm on the average, thus having no consequences for the predicted trends.\\

\end{document}